# Predicting Software Effort from Use Case Points: A Systematic Review

Mohammad Azzeh
Department of Software Engineering
Applied Science University
Amman, Jordan POBOX 166
m.y.azzeh@asu.edu.jo

Ali Bou Nassif
Department of Computer Engineering
University of Sharjah
Sharjah, UAE
anassif@sharjah.ac.ae

Imtinan Basem Attili
Department of Electrical Engineering
University of Sharjah
Sharjah, UAE
iattili@sharjah.ac.ae

Abstract.

*Context*: Predicting software project effort from Use Case Points (UCP) method is increasingly used among researchers and practitioners. However, unlike other effort estimation domains, this area of interest has not been systematically reviewed.

*Aims*: There is a need for a systemic literature review to provide directions and supports for this research area of effort estimation. Specifically, the objective of this study is twofold: 1) to classify UCP effort estimation papers based on four criteria: contribution type, research approach, dataset type and techniques used with UCP; and 2) to analyze these papers from different views: estimation accuracy, favorable estimation context and impact of combined techniques on the accuracy of UCP.

*Method*: We used the systematic literature review methodology proposed by Kitchenham and Charters. This includes searching for the most relevant papers, selecting quality papers, extracting data and drawing results.

*Result*: The authors of UCP research paper, are generally not aware of previous published results and conclusions in the field of UCP effort estimation. There is a lack of UCP related publications in the top software engineering journals. This makes a conclusion that such papers are not useful for the community. Furthermore, most articles used small numbers of projects which cannot support generalizing the conclusion in most cases.

*Conclusions*: There are multiple research directions for UCP method that have not been examined so far such as validating the algebraic construction of UCP based on industrial data. Also, there is a need for standard automated tools that govern the process of translating use case diagram into its corresponding UCP metrics. Although there is an increase interest among researchers to collect industrial data and build effort prediction models based on machine learning methods, the quality of data is still subject to debate.

## 1. Introduction

Use Case Points (UCP) has been studied intensively in the past two decades as an alternative method to Function Points, with a goal to predicting software effort at early software development phases [1][2][3]. In spite of the widespread of use case techniques in software industry, the UCP method still confronts some challenges such as: (1) lack of sufficient public industrial data [4], (2) different human interpretation [5], (3) the debate about algebraic construction of UCP method [1], (4) the way that the use case complexity is measured, and most importantly (5) there is no stable technique in literature that enables practitioners to produce most likely effort from UCP size metrics. remarkably, not all published research studies on UCP provide significant contributions to body of knowledge of the research field. In fact, a good number of studies attempt to only run examples on UCP method without supporting the community with useful conclusions. Therefore, there is a need for a Systematic Literature Review (SLR) that can help in classifying and analyzing the published studies with a goal to structure this field of interest and provide directions and support for future research [6][7]. Despite many systematic literature reviews on effort estimation, the field of effort estimation based on UCP is still not reviewed carefully. It is well known that the goal of SLRs is not only aggregate all existing evidence on research questions, but also to support the development of evidence-based guidelines for practitioners [6][8][9].

This paper introduces a systematic literature review for UCP method. This allows us to: 1) build a classification scheme and structuring the field of interest, and 2) provide recommendations based on the strength of the evidence of UCP effort estimation performance in the current research. In fact, we reviewed over 120 published research studies on effort estimation based on UCP. These studies have been identified from different research venues in the period from 1993 to February 2020. We chose the year of 1993 because the original model of UCP effort estimation was proposed in that year by Karner [10]. In order to ensure the quality of these studies and their benefit for our systematic mapping and literature review, we followed the procedure proposed by Kitchenham and Charters [11] based on evidence-based software engineering. Basically, we identified some mapping and review questions in order to classify the published UCP effort estimation studies based on various criteria. To facilitate the searching process, we constructed a comprehensive search query to find candidate papers. Then, inclusion and exclusion criteria are used to filter the most relevant papers, in addition, the quality assessment is run on the selected papers in order to filter the most quality papers that help us to draw better conclusion. lastly, the data extraction, analysis and synthesis are performed to extract knowledge and conclusions. At the end of these steps we ended up with 75 quality papers that help us to address mapping and review questions. Our paper is different from previous SLRs in the following points:

1. *Different focus*: The focus of this study is on a narrower field of effort estimation which is based on UCP size metric only. The UCP method itself has not be reviewed thoroughly in previous SLRs as a mean of effort estimation.
2. *Structuring the field of interest*: this study aims to build classification scheme based on effort estimation and UCP. The classification scheme includes classifying and studying UCP papers based on their contribution, favorable prediction techniques, research approaches and nature of datasets.

The rest of the paper is organized as follows: Section 2 presents related work on previous SLR studies in the field of software effort estimation. Section 3 introduces the method of SLR that we used in this paper. Sections 4 and 5 present results of mapping and review questions. Sections 6 and 7 present discussion on the findings and conclusions.

## 2. Related Work

When researchers encounter software engineering literature, they found multiple systematic review studies on effort estimation, but just one of them was designed for UCP effort estimation. The reason behind that might be because they involve UCP effort estimation methods with other effort estimation methods. A recent systematic literature review study was conducted by Mahmood et al. [12] based on UCP and experts judgement software effort estimation methods. They identified only 34 papers within the period 2000 to 2019. The main theme of this study is to examine the published paper on UCP and expert judgement in terms of several point of views including research contribution, dataset usage, accuracy metrics, and findings of the selected studies. They found that expert judgment estimation technique is the most frequently used in effort estimation. In addition to that, the industrial datasets are the most used type of data in UCP method.

The main difference between our SLR study and Mahmood et al. [12] study are:

1. Different theme, they focus on two estimation approaches which are Use Case Points and Expert Judgement. Among 34 studies, only 25 studies are related to UCP. Therefore, great number of valuable publications in the domain has been ignored.
2. Our SLR is a mapping and review studies, while their study is just review.
3. Search string, inclusion and exclusion criteria and research questions are quietly different.
4. Mahmood et al. [12] focused on public datasets that are not related to UCP, whereas we focused our study on UCP public and private datasets.

On the other side, we present a literature for common SLR studies in software effort estimation. Jørgensen and Shepperd [9] conducted a systematic literature review on software development cost estimation studies. The purpose of this SLR was to increase awareness of researchers with quality and reliability of previous published studies. They identified 304 relevant conference and journal papers published in different venues. The outcomes of this SLR provided recommendations about identifying relevant research papers and selecting appropriate research venues. They also recommended using industrial software effort estimation methods and datasets.

Idri et al. [7] conducted a systematic mapping and literature review on analogy-based effort estimation. They used a comprehensive SLR process to cover all aspects of analogy-based research area including estimation accuracy and context. The results of this SLR focused the light on the current research directions on this kind of estimation and open new research questions on how to improve the visibility of analogy-based effort estimation.

Usman et al. [13] conducted a systematic literature review on software effort estimation in agile software development. They selected 32 quality papers out of 443 candidate papers. Among various estimation techniques, they found expert judgment, planning poker, and use case points method are the most used estimation techniques. However, none of them achieved good accuracy. Furthermore, there was a debate about which reliable cost drivers that should be applied for this kind of estimation. The story points and use case points were the most used size metric in agile software development.

Kitchenham et al. [6] investigated and evaluated the previous SLR studies in software engineering. They conducted a systematic review and showed that eight out of twenty relevant studies addressed research trends rather than technique evaluation, and seven SLRs addressed cost estimation. The study suggests the following: 1) mainstream software engineering topics are not well represented. 2) the Simula Research Laboratory in Norway is the leading software engineering institution in terms of undertaking SLRs. 3) the current output of Evidence Based Software Engineering (EBSE) articles is strongly supported by European researchers.

In the same direction MacDonell et al. [14], investigated the reliability of previous systematic reviews in the context of empirical software engineering. They conducted two parallel reviews of the same research questions based on an agreed metaprotocol with a goal to measure the outcome stability of systematic reviews. Both teams worked separately in parallel. The results found that there is no significant difference between two outcomes, which showed that the SLR proved to be robust to differences in process and produced stable outcomes.

## 3. Method

The procedure of SLR that we used in this study includes five main steps, as shown in Figure 1. In the first step, we identified main mapping and review questions. The goal of mapping review is to enable us classifying the published UCP effort estimation studies based on various criteria. In the second step, we construct the search query that will be used to find candidate papers on the common digital libraries such as IEEE Xplore, ACM, Science Direct, Wiley and Springer Link. In the third step, the inclusion and exclusion criteria are applied on all candidate papers to filter the most relevant papers. During this step, the duplicate papers are also removed. Later, the quality assessment is run on the selected papers from the previous step in order to filter the most quality papers that help us to draw better conclusion and extract required knowledge. In the last steps, the data acquisition, analysis and synthesis are performed in order to extract knowledge and findings. In the upcoming subsections the complete description of each step is discussed in more details.

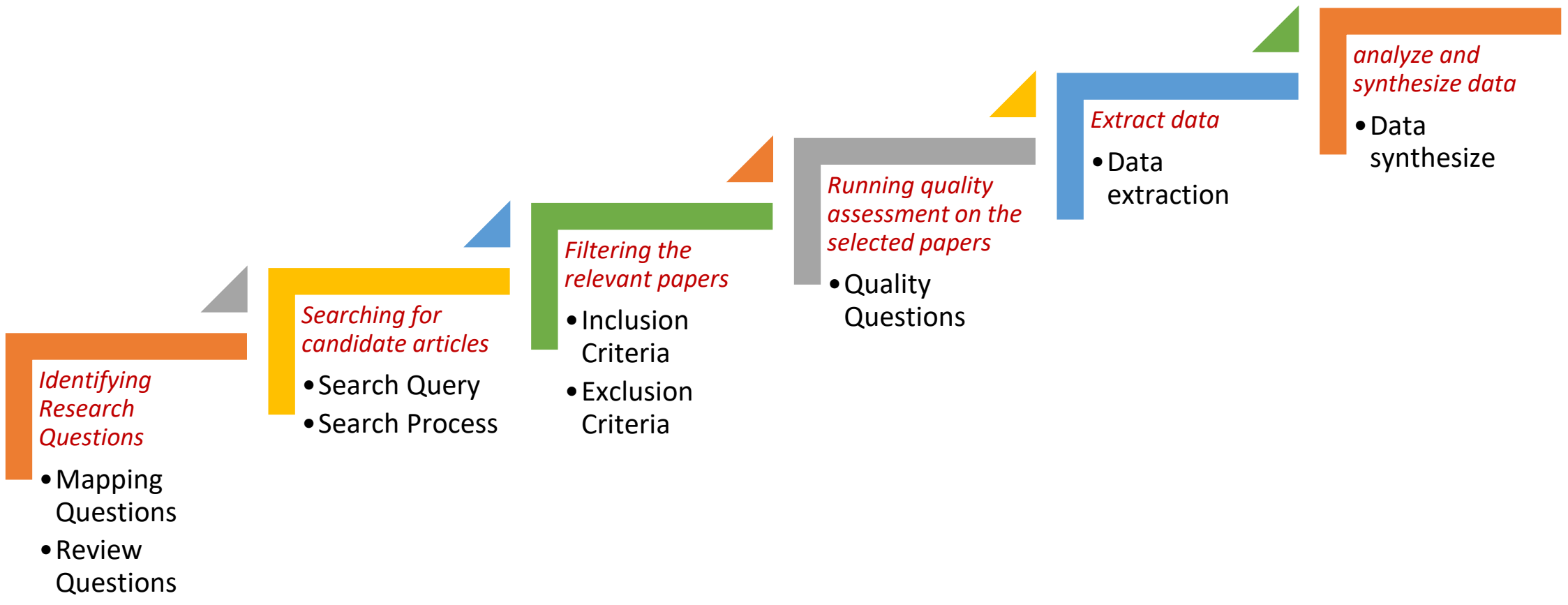

Fig. 1 Mapping and Review Process

### 3.1 Research Questions

Nine research questions (5 mapping questions and 4 review questions) were proposed to achieve the goals of this SLR study. The obtained review results enable researchers to understand the main types of contributions and challenges in this field of interest. The synthesized data can usually serve as basis for recommendations of the best practices. The Nine research questions have been proposed and reviewed by three authors in order to avoid inconsistency among them. At glance, these questions are classified into two categories: mapping and review questions. The mapping questions (MQ1 to MQ5) aim at building a classification scheme and structuring the field of interest. The classification criteria that we used are described in Table 1. The review questions (RQ1 to RQ4) aim at providing recommendations based on the strength of the evidence of UCP effort estimation performance in current research. The mapping and review questions are summarized as follows:

- Mapping Questions:
  - MQ1. Which and how many sources include papers on effort estimation based on UCP?
  - MQ2. What are the main contributions of UCP research papers?
  - MQ3. What are the main types of research approaches applied to effort estimation based on UCP?
  - MQ4. What are the most favorable techniques that are used to produce effort estimates from UCP, and how has this changed over time?
  - MQ5. What are the types of datasets that are used to run experiments for UCP effort estimation?
- Review Questions:
  - RQ1. How easy is to collect UCP data from software industry and what is the implication on effort estimation?
  - RQ2. What are the impacts of combining other techniques with an UCP method technique on its estimation accuracy?
  - RQ3. What is the most favorable context for UCP effort estimation method?
  - RQ4. To what extent are researchers on UCP methods aware of the breadth of potential estimation study sources?

Table 1. Classification Criteria

| Property | Categories |
| --- | --- |
| Contribution Type | technique, tool, comparison, validation, Metric and model |
| Research Approaches | Proposal, Evaluation, Review, Case Study, Survey, Theory, Comparative studies |
| Techniques used in combination with UCP | Fuzzy Logic Neural Networks, Multiple Linear Regression, Support Vector Regression, Stepwise Regression |
| Dataset Type | Industrial, Educational, Mix of Both Industrial and Educational, and Case studies. |

The purpose of MQ1 is to support researchers with the most relevant studies in UCP effort estimation. To address MQ1, we have reviewed the common venues of effort estimation publications such as "IEEE Transactions on Software Engineering, ACM Transactions on Software Engineering and Methodology, Information and Software Technology, Journal of Systems and Software, Empirical Software Engineering." Also, some papers published in different journals or conference proceedings have been reviewed. MQ2 aims to list various types of contributions in UCP. We identified the possible main contributions in the field of interest. These contributions are classified to 1) technique, 2) tool, 3) comparison, 4) validation, 5) metric and 6) model as shown and described in Table 2. MQ3 attempts to identify the potential categories of applied research approaches for UCP effort estimation studies. We have identified seven research categories that have been applied in this field of interest as shown and described in Table 3.

Table 2. Types of contributions

| Contribution Type | Description |
|---|---|
| Technique | A new UCP technique is proposed, or existing one is improved. |
| Tool | Proposing a new tool for UCP effort estimation |
| Comparison | A comparison among different configurations of UCP, or comparison with other techniques. |
| Validation | An evaluation of the performance of existing UCP techniques using some historical datasets. |
| Enhancement | An enhancement made to existing or original UCP model. |
| Model | A new UCP effort estimation model is developed using machine learning or data mining methods. |

Table 3. Types of research approaches

| Research Approach | Explanation |
|---|---|
| Proposal | A new model or improved model for UCP. |
| Evaluation | Evaluating existing UCP or effort estimation models based UCP. |
| Review | Studies that reviewed UCP or Effort estimation based on UCP. |
| Case Study | Empirical evaluation of UCP and effort estimation based on case studies. |
| Survey | Studies that provide comprehensive survey for UCP. |
| Theory | Studies that evaluate UCP structure or part of its properties theoretically, not empirically. |
| Comparative studies | Studies that provide a comparison among different effort estimation models based on UCP. |

To address MQ4, we identify the most prediction techniques that are used to generate effort estimates from UCP. We have identified all possible methods that were used to build effort estimation model based on UCP such as: Karner (i.e. original model of UCP) [10], Fuzzy Logic [3], [15], Neural Networks [17][18], Multiple Linear Regression [18]. Finally, MQ5 is designed to investigate the type of data the was used to validate and check accuracy of various UCP effort estimation models. The datasets types were classified into four categories: 1) Industrial, 2) Educational, 3) Mix of Both Industrial and Educational, and finally 4) Case studies.

With respect to RQ1, we discuss the impact of used data and validation technique on the accuracy of UCP effort estimation. We discuss the accuracy for different types of used datasets. For RQ2, we identify the impact of using other machine learning and data mining techniques on the accuracy of UCP method. RQ3 aims to identify the different UCP research contexts that are most favorable among researchers. we identify the challenges that appear when using student projects and case studies for building effort estimation based UCP models. we consider the assessment of the vulnerability of UCP computing procedure (using transaction points or steps) and weights on effort estimation. Finally, RQ4 intends to shade the light on the possible shortcomings when searching for relevant work.

### 3.2 Search process

To address the proposed research questions, we made an automated search query, using predefined searching terms, in the common research digital libraries and electronic databases such as "IEEE Digital Library, ACM Digital Library, Science Direct, Wiley, Springer". These databases were selected because, from experience, most of the journals and proceeding of the selected papers were indexed by these databases. Furthermore, Google Scholar was also used to search for papers that were not indexed in the above libraries because it explores other digital libraries on the Internet.

With respect to query string formulation, we followed the SLR guidelines mentioned in [11]. The query string is constructed using all important terms then these terms are linked using AND/OR operators. OR operator is used to link all synonyms and variation of each incorporated terms, then the AND operator is used to connect the main terms. All authors have participated in the query construction and revision to ensure its comprehensive and correctness. The formulated query string has been tested in some digital libraries to ensure its quality. The initial query string did not include terms such as Lifecyle and Simplifying which they have been added after testing on some digital libraries such as IEEE digital library and Science Direct. The query string has been executed in February 2020. Keywords from other known studies (e.g. title, abstract and keywords from well-known studies) have been added if they were not part of the initial query string. The complete proposed query string is:

> *("Use Case Points" OR "Use Cases" OR UCP) AND (Project OR Software OR System OR Application OR Product OR Development OR Lifecycle) AND (Effort OR Cost) AND (Prediction OR Predicting OR Estimation OR Estimating OR Evaluating OR Simplifying*)

In the initial search stage, we used the proposed query string to search for candidate papers in the mentioned digital libraries. In the second search stage, we have reviewed the reference lists of the filtered studies that met inclusion and exclusion criteria (see section 3.3) to identify papers related to UCP based on their title. The papers that were found highly relevant are then added to the list of relevant papers. Furthermore, some papers that we already aware of have been used to control the quality of the search as shown in Table 4. We recorded for each control paper the digital library from which it was retrieved before and after the search. Only four cases present mismatch result because of the sequence of digital library search (IEEE, ACM, Springer, Science Direct, Wiley and Google Scholar). This enables us to assess whether or not the initial search stage had missed any highly relevant papers and to ensure that the search covered the maximum number of available UCP studies.

According to Kitchenham et al. [11], we also formulated the query string elements using Population, Intervention, Comparison, Outcome, Context) PCIOC approach:

Population: software project.
Intervention: Use Case Points effort estimation model.
Comparison: Variants of Use Case Points effort estimation model
Outcomes: Prediction or estimate accuracy.

Table 4. List of known existing papers used to validate the search string.

| Paper Id | Database before search | Data based after search |
|---|---|---|
| S5 | IEEE | IEEE |
| S6 | IEEE | IEEE |
| S8 | ACM | ACM |
| S9 | ACM | ACM |
| S20 | Google Scholar | IEEE |
| S24 | Google Scholar | Google Scholar |
| S25 | Google Scholar | Google Scholar |
| S35 | Science Direct | Science Direct |
| S36 | Google Scholar | IEEE |
| S41 | Science Direct | Science Direct |
| S46 | ACM | ACM |
| S56 | Google Scholar | Science Direct |
| S62 | Wiley | Wiley |
| S63 | Science Direct | Science Direct |
| S68 | IET Software | IET Software |
| S69 | Wiley | Wiley |
| S70 | Science Direct | Science Direct |
| S73 | Google Scholar | IEEE |

### 3.3 Inclusion and Exclusion Criteria

This process is conducted to filter the selected papers from previous step by running inclusion and exclusion criteria. This step is important as it enables us to identify the relevant and related papers that can help in answering research questions. First, all authors discussed the criteria of inclusion and exclusion that should be used. After reaching an agreement on the inclusion and exclusion criteria, each paper (mainly title, abstract and keywords) was evaluated by first and second authors based on the recommended inclusion/exclusion criteria to determine whether it should be retained or rejected. The paper is considered "*include*" if at least meets one of inclusion criteria and none of exclusion criteria. On the other hand, the paper is considered "*exclude*" if at least one of the exclusion criteria and none of inclusion criteria is met. Finally, the paper is considered "uncertain" if it meets some inclusion and exclusion criteria. In this case if all authors agreed that the paper is "*include*" then the paper is retained, otherwise the paper is considered "*exclude*", thus it is rejected. If the decision could not be made, all authors discussed the paper again by reading the full text until they reach agreement. The high number of agreements on the decision confirm the relevance of the proposed inclusion and exclusion criteria. The following points explain the inclusion and exclusion criteria in which OR Boolean operator is used to link between points in each type.

Inclusion criteria:

- Using the UCP to predict effort estimation, OR
- Comparison between different effort estimation based on UCP, OR
- improving UCP sizing method.

Exclusion criteria:

- Using UCP for late software project development such as testing and maintenance, OR
- Duplicate studies (only complete version is selected), OR
- UCP description and explanation papers, OR
- Comparing between UCP and other software size metrics (e.g. Function Points, Object Points), OR
- White papers and Newsletters about UCP.

### 3.4 Quality assessment

Quality assessment is an important step in SLR to ensure the quality of the relevant studies. Also, to limit bias in conducting SLR. This process comes after inclusion and exclusion step. The questions that we used for quality assessment are:

QA1. Are the objectives of the study clear?
QA2. Does the study add value to the existing literature?
QA3. Is the methodology of the experiments clearly described?
QA4. Is the used dataset suitable for this type of studies?
QA5. Are the findings supported by performance and statistical results?
QA6. Are the threats to validity mentioned?
QA7. Was the study published in high-quality journal or conference proceedings?

The third author ran quality assessment on the selected papers independently. She reviewed each paper and gave score for each question based on the criteria discussed below. When the decision could not be reached, the third author communicated with other authors to give their opinion and to reach an agreement. The total score for each paper is then computed by summing the questions scores. If the total score is less than or equal to 4, the paper is excluded at this stage, otherwise it is retained. The questions were scored as follows:
QA1: score (+1) is given if the objectives are stated clearly. Score (+0.5) is given if the objectives are not well defined. Score (0) is given if the objectives are not defined at all.
QA2: score (+1) is given if the study contributed noticeably to the existing literature. (+0.5) is given if the study contributed partially to the literature. (0) is given if no contributions were added to the literature.
QA3: score (+1) is given if the methodology of research is clearly described including descriptions of validation strategy, dataset and performance measures. (+0.5) is given if the methodology described part of the previous components. (0) is given if no methodology description is given in the paper.
QA4: score (+1) is given if the size of the dataset is sufficient enough to support the findings and quality of projects is accepted. (+0.5) is given if the dataset size is not large or quality of data is partially met. (0) is given if there is no dataset to support the findings.

QA5: score (+1) is given if there are results and statistical tests to support the findings of the study. (+0.5) is given if there are results but there is no statistical test to support the findings. (0) is given if the results are weak, and no statistical tests were conducted.
QA6: score (+1) is given if the paper reports the threats to validity in a correct way. (+0.5) is given if the threats to validity are mentioned but not related to methodology. (0) is given if no threats to validity was given.
QA7: to answer this question we use on the Journal Citation Report (JCR) 2019 or Science Citation Index Expanded (SCIE) to score the journal of the paper, and the Computer Science Conference Ranking (CORE) to score conference/workshop/symposium of the paper. Score (+1) is given if the journal is ranked Q1 or Q2 in JCR or indexed in SCIE, or if the conference/workshop/symposium is ranked A in CORE. (+0.5) is given if the journal is ranked Q3 in JCR, or if the conference/workshop/symposium is ranked B in CORE. (0) is given if the journal is ranked Q4 in JCR, or if the conference/workshop/symposium is ranked C in CORE.

### 3.5 Data extraction and synthesize

As recommended in the SLR guideline [11], The data extraction was carried out by the first two authors of this paper, where both authors extracted and checked the data. However, the first author coordinated the extraction and checking data procedure. Each paper was read by the three authors and the necessary data were collected with goals to serve objectives of our SLR. Any disagreement among the authors regarding a particular paper was resolved by contacting the authors of that paper to answer some necessary questions about their paper. The list below contains data extraction form which describes the data that have been extracted from candidate papers. These data are then summarized and tabulated in a way that can serve the objectives of our SLR. The data that we have extracted are:

- Article Identifier.
- Article Title.
- The articles source.
- Publication Venue.
- Publication year
- Authors
- Objectives.
- Quality evaluation.
- Techniques that are used to construct the effort estimation.
- Research type.
- Contribution type.
- Datasets used.
- Keywords.

After that, the summarized data are synthesized in order to answer the proposed research questions. Since the data contain quantitative and qualitative data, different kinds of synthesis approaches were used such narrative synthesis and quantitative data synthesis to show the synthesized data in different forms. However, approaches like meta-analysis is not possible because the studies included in our SLR are not similar due to different experiments configuration and different datasets.

### 3.6 Threats to Validity

The main threat to validity of our review is the exclusion of relevant articles. Finding the most relevant and quality papers is the main task of an SLR study because it has significant impact on the findings and the drawn conclusions. To reduce this threat, all authors participated in designing the search string. The proposed search string was evaluated on the electronic digital libraries in order to examine its accuracy in retrieving related papers. The first two authors checked manually the references of the selected papers separately, in parallel, in order to identify all missing articles that were not returned by the search string. Furthermore, two authors separately conducted applying inclusion/exclusion criteria and quality assessment, if there was any doubt, the full study was read again. All disagreements among authors were discussed until a final consensus was reached. For quality assessment, the minimum threshold was set by all authors to be 4 out of 7 which enables us to select the most relevant papers. The second threat is publication bias. Basically, we have identified five main electronic digital libraries that are commonly have strong publications in the field. Google scholar was also used to apply the search string because it explores other digital libraries on the internet. Google scholar allows us to not avoid publications venues that are not commonly used among software engineering researchers. The third threat is data extraction bias. Two authors conducted exhaustive search and read each paper independently to extract important data from the selected papers. We have identified the primary data that helped us to address the research questions as explained in section 3.5. However, data extraction bias may occur, especially when collecting data from papers that present case studies or that used unclear dataset. Therefore, the data extracted for each paper were compared and all disagreements were discussed by the researchers.

## 4. Results of Mapping Questions

This section describes the results of mapping questions (MQs). Firstly, we made search for candidate papers using the proposed search string on the common digital libraries (IEEE Xplore, ACM, Science Direct, Wiley, Springer Link and IET). We also applied the search query on google scholar because it is a common indexing venue for other digital libraries. This process has resulted with total of 127 related papers as shown in Figure 2, most of them were collected from IEEE Xplore digital library. After that, the inclusion and exclusion criteria have been applied to filter out all the relevant papers. This step has resulted 91 (72%) selected papers. Finally, the quality assessment has been applied on the selected papers from previous step by assessing and aggregating quality scores for each paper as shown in Appendix B. This step has resulted 75 (59%) acceptable quality papers where 33.3% of them (i.e. 25 out of 75) were high and very high quality as shown in Table 5.

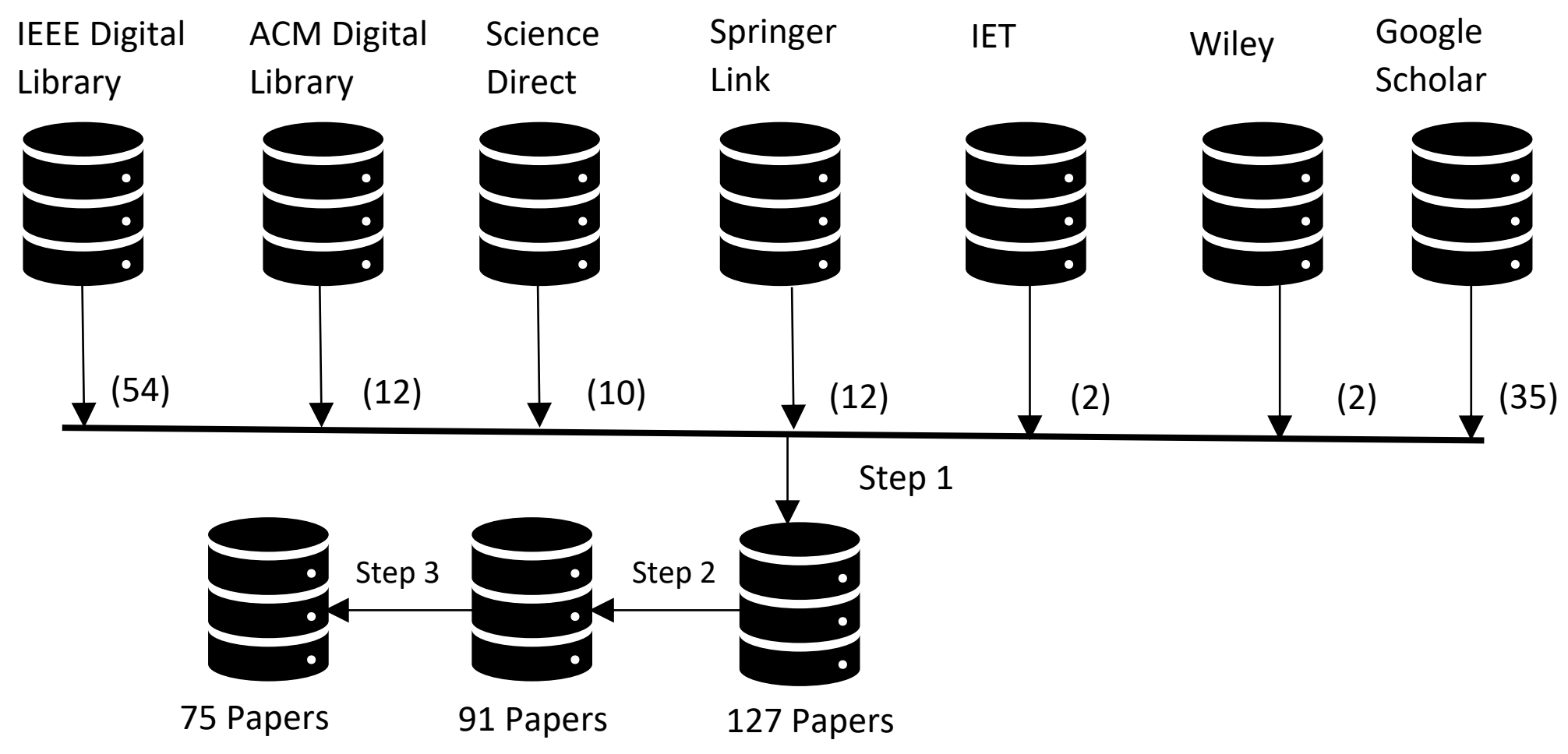

Fig. 2 Results of searching process

The complete lists of the selected studies with their citations and classification are presented in Appendixes A and C. Figure 3 shows the number of papers in each searching phase classified as conference paper, journal article, magazine, newsletter and white paper. It is clearly noted that most of the papers have been published in conferences and few of them were published in journals. This may explain why this type of effort estimation is not yet mature enough like other estimation disciplines. We kept only one white paper, which is the Karner paper [10], because it describes the original UCP model, which is the basis of effort estimation based UCP. Figure 4 shows the publications over the years, colored with type of publications. Clearly, there was an increasing interest in effort estimation based on UCP among researchers in the beginning of 2000s and reached peak in 2005, then it began to decline with some attempts to stabilize the publications. Notably, the journal articles started to appear after 2005 which demonstrates that this area of research become mature after 2005. Number of publications for this kind of estimation is still small in comparison to other type of effort estimation. Figure 5 shows that the majority of selected papers have been found on IEEE Xplore digital library because the common conferences that authors usually published were sponsored or technically co-sponsored by IEEE. This is because we executed the query string first on the main research digital libraries such as IEEE, ACM, Science Direct, Springer Link and the papers found in these libraries are counted for them. Then the remaining papers that have been found in Google scholar are counted for Google scholar. this explain why IEEE digital library has larger number of publications than Google scholar.

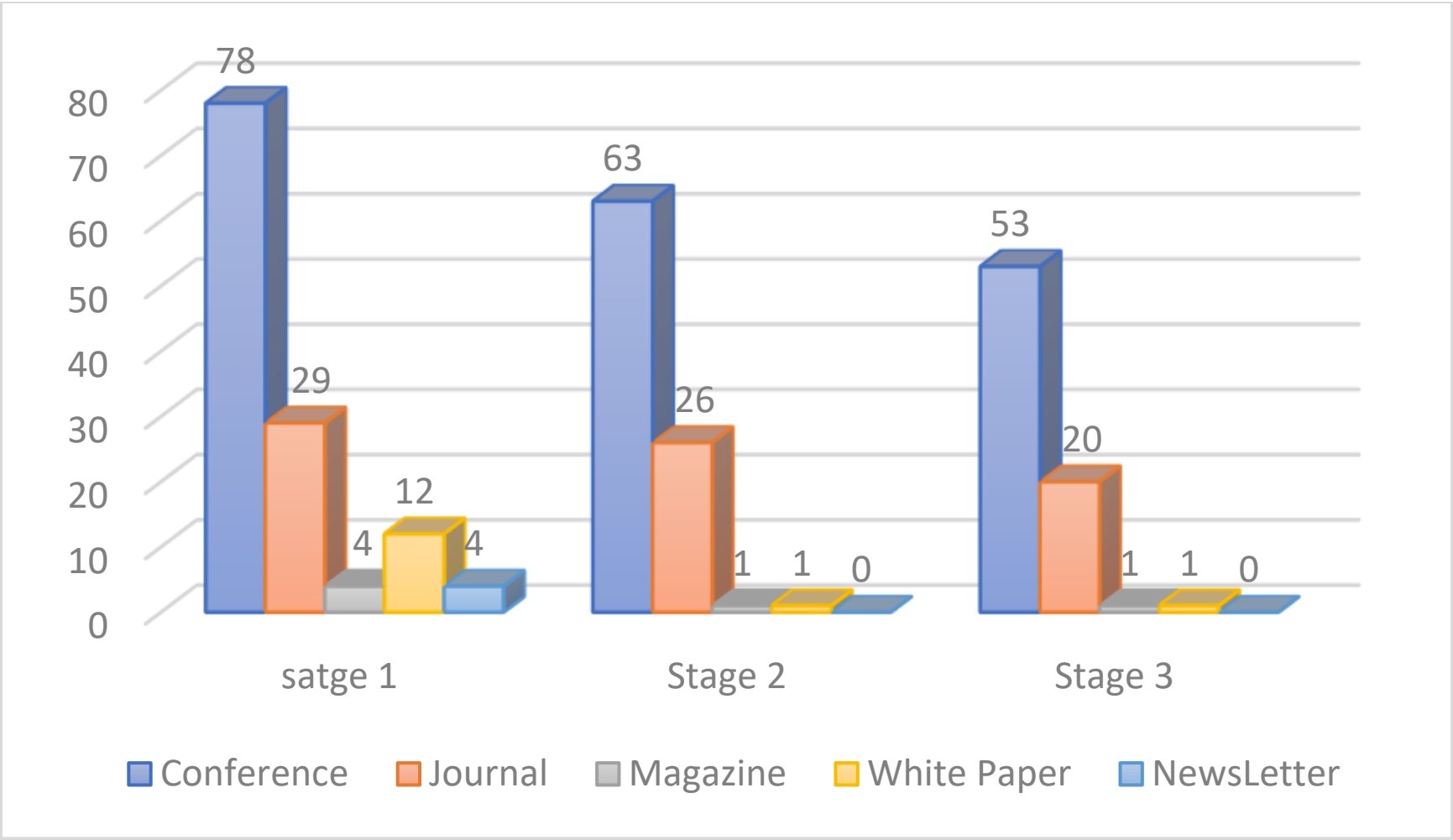

Fig3. Number of papers after each stage of searching and filtering process

Table 5. Quality levels of the selected studies.

| Quality Level | No. of Studies | %proportion |
|---|---|---|
| Very high (6<quality score ≤ 7) | 11 | 0.121 |
| High (5<quality score ≤6) | 14 | 0.154 |
| Average (4<quality score ≤5) | 50 | 0.549 |
| Low (3<quality score ≤4) | 9 | 0.099 |
| Very Low (0<quality score ≤3) | 7 | 0.077 |
| Total | 91 | 1.0 |

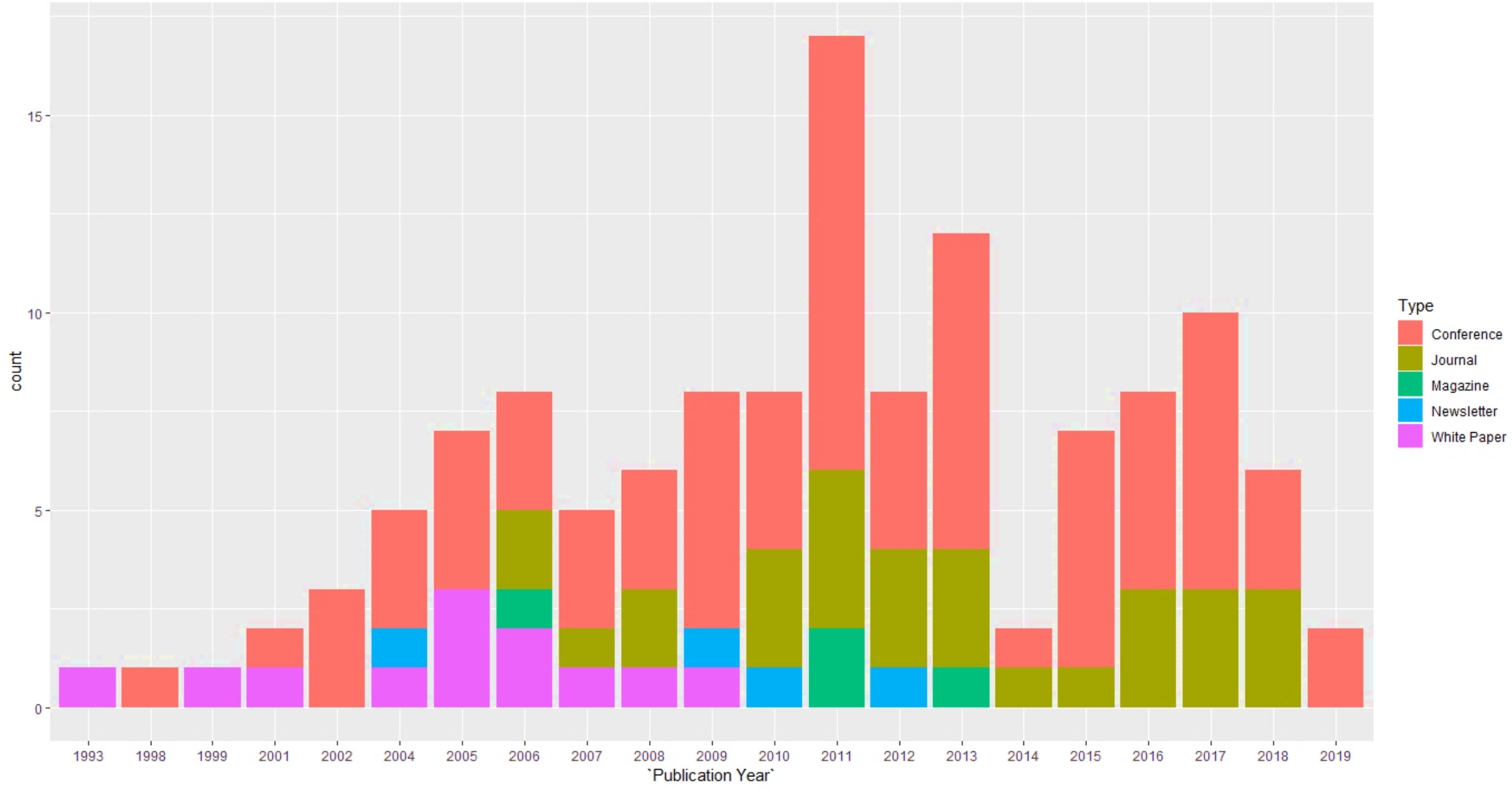

Fig 4. Publications over the years, based on the article type.

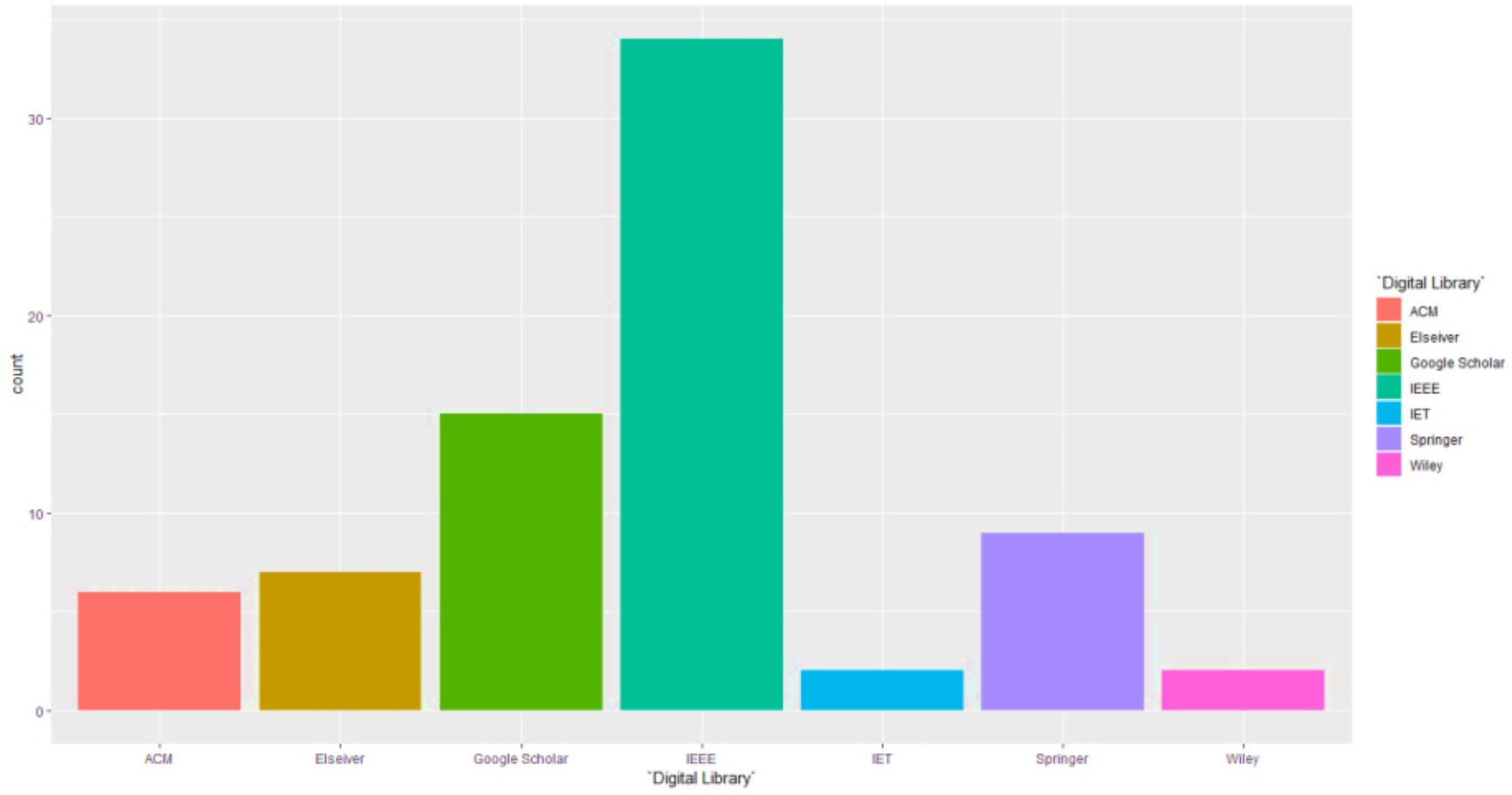

Fig 5. Number of papers in every digital library

## 4.1 Venues of UCP Effort Estimation Publications (MQ1)

As mentioned earlier we identified 75 relevant papers after two stages of filtering. Most of these papers were published in conferences and just few of them were published in journals as shown in Figure 1. The top journals that have papers on UCP effort estimation are summarized in Table 6. We can notice that there are no common venues among publications. Also, not all listed journals are classified as software engineering research focus. Only 7 out of 14 venues can be considered software engineering journals, which have 13 out of 20 publications. The remaining seven publications were published in none-focused software engineering journals which dramatically affect the visibility and breadth of the quality of forthcoming papers. This raises a big concern about the reputation of the published work in this field of interest. Therefore, the outcome of those papers that were published in none-focused software engineering journals would be biased toward the theme of that journal and their findings would seem unuseful for software engineering community. Unsurprisingly, the top five journals in the table that have two publications or more are classified as software engineering journals, in addition to the journal of International Journal of Software Engineering and Knowledge Engineering and The Journal of Defense Software Engineering that have one publication. Indeed, reading only the top 5 most relevant journals means that important research results may be missed. The other top ranked journal in the field of software engineering such as IEEE transactions on Software Engineering, Empirical software engineering, ACM transactions on software engineering and methodology are not yet having papers on UCP effort estimation.

Table 6. Most Important UCP Effort Estimation Journals

| "Journal Name | # Publications | Reference |
|---|---|---|
| Information Software and Technology | 3 | [S13] [S35][S70] |
| Journal of Systems and Software | 2 | [S41][S63] |
| Innovations in Systems and Software Engineering | 2 | [S15][S64] |
| IET Software | 2 | [S57][S68] |
| Journal of Software: Evolution and Process | 2 | [S62][S69] |
| Applied Soft Computing | 1 | [S56] |
| International Journal of Information Processing | 1 | [S51] |
| International Journal of Computer Applications | 1 | [S47] |
| International Journal of Intelligent Systems and Applications | 1 | [S40] |
| International Journal of Software Engineering and Knowledge Engineering | 1 | [S34] |
| Foundations of Computing and Decision Sciences | 1 | [S25] |
| Journal of Global Research in Computer Science | 1 | [S24] |
| INFOCOMP Journal of Computer Science | 1 | [S16] |
| The Journal of Defense Software Engineering | 1 | [S11]" |

### 4.2 Main contributions (MQ2)

We have been interested in getting to know what are the main contribution types in this field of interest. We have identified and described six possible research contributions as shown in Table 2. The classification of papers based on contribution types is presented in Table 7. After studying and analyzing the selected papers by three authors, we found that that Enhancement contribution type is the most dominant one, followed by validation contribution type with very slight difference as shown in Figure 6. This is an indication that most authors favor either continue developing new models on the original UCP sizing method or validating existing one on industrial and new datasets from different domains. The third top ranked contribution type is developing techniques to estimate effort from UCP size metrics. Building tools to measure UCP from use case diagram and requirements are very rare and have little of interest among researchers. From these results we can understand that researchers do not interest in developing effort estimation models based on UCP as much as their interest in enhancing the sizing mechanism of UCP method.

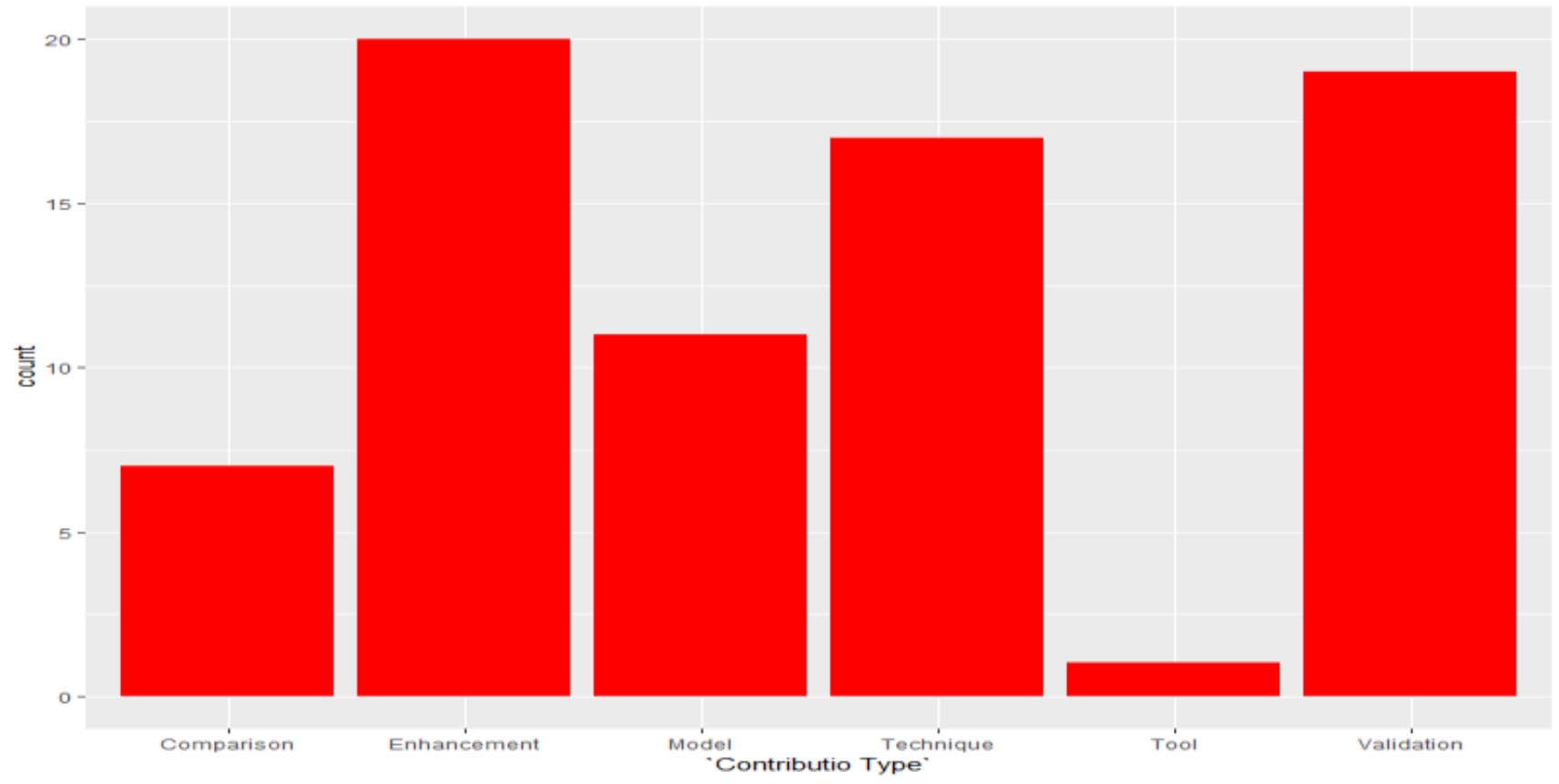

Figure 6. Bar plot for main contributions types

Table 7. Studies classification based on contribution type.

| “Contribution Type | Studies | Proportion |
|---|---|---|
| Technique | [S1][S12][S15][S16][S21][S24][S25][S26][S29][S30][S32][S34][S45][S49][S54][S63][S71][S73] | 24% |
| Tool | [S6] | 1.3% |
| Comparison | [S3][S10][S22][S27][S44][S58][S68] | 9.3% |
| Validation | [S2][S7][S9][S11][S23][S33][S35][S43][S59][S62][S64][S65][S66][S67][S69][S70][S74][S75] | 24% |
| Enhancement | [S4][S5][S8][S14][S17][S19][S28][S31][S37][S42][46][S47][S48][S50] [S52][S53][S55][S60][S61][S72] | 26.7% |
| Model | [S13][S18][S20][S36][S38][S39][S40][S41][S51][S56][S57] | 14.7%” |

### 4.3 Main research approaches (MQ3)

The third proposed research questions stated that:” *What are the main types of research approaches applied to effort estimation based on UCP?*”. Indeed, we are interested in getting know which are the main research methodologies that were used by researchers when studying this field of UCP effort estimation. The classification of papers based on research approach is shown in Table 8 and Figure 7. We can notice that the most studies focus on evaluating UCP effort estimation as potential approach for early effort estimation. The second approach is the case study. This approach was used by around 20 studies to examine the usefulness of UCP effort estimation based on specific case studies. The third approach is based on comparison between different effort estimation models based on UCP such as comparing between Fuzzy Logic and Neural network over UCP historical projects.

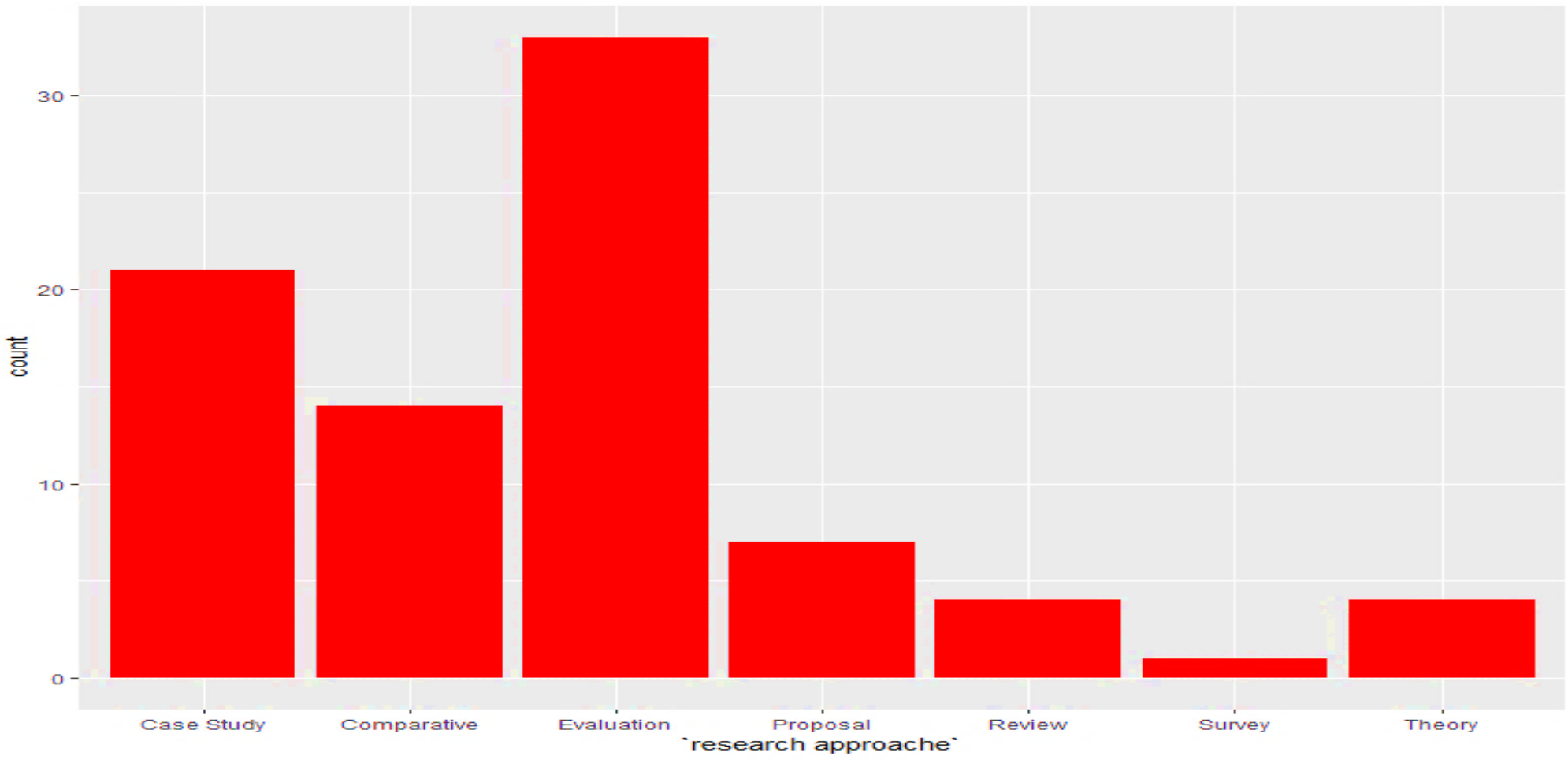

Fig7. Bar plot for main research approaches.

Table 8. Studies classification based on research approach.

| “Research approach | Explanation | Proportion |
|---|---|---|
| Proposal | [S1][S37][S48][S49][S58][S59][S72] | 9.3% |
| Evaluation | [S5][S12][S14][S16][S19][S21][S23][S24][S26][S29][S30][S32][S33][S35][S36][S38][S39][S40][S41][S45][S46][S47][S54][S55][S57][S60][S61][S63][S64][S65][S67][S69][S70] | 44% |
| Review | [S34][S43][S66][S74] | 5.3% |
| Case Study | [S1][S2][S4][S6][S7][S8][S9][S10][S11][S15][S18][S19][S20] [S25][S49][S53][S62][S65][S71][S72][S75] | 28% |
| Survey | [S34] | 1.3% |
| Theory | [S17][S28][S31][S42] | 5.3% |
| Comparative studies | [S3][S12][S13][S22][S27][S44][S46][S50][S51][S52][S54][S56] [S57][S68][S73] | 20%” |

### 4.4 MQ4. What are the most favorable techniques that are used to produce effort estimates from UCP, and how has this changed over time?

The principal problem in UCP method is how to convert the likely UCP size metric into its corresponding project effort. The early version of UCP method that is proposed by Karner [10] suggested using productivity as second cost driver bedside UCP, where a fixed Productivity or a very limited productivity ratios were used [3], [19], [20]. These productivity values were figured out based on a very limited number of case studies. This approach was the dominate approach when estimating effort from UCP in most previous papers. The main reason behind using this approach was due to the absence of large historical industrial data, so authors used the original model to validate their hypothesis. In addition to that basic approach, authors attempt to build machine learning or statistical models to estimate effort from UCP based on small number of projects. Figure 8 depicts the most common techniques that were frequently used in combination of UCP in order to improve its accuracy. Amongst them, fuzzy logic, multiple linear regression and neural network have been the dominant methods used to estimate effort from UCP size metrics. The main problem with machine learning is that they need sufficient data in order to build a reliable prediction model. We can also notice that the most techniques that were published in journals are Karner and multiple linear regression.

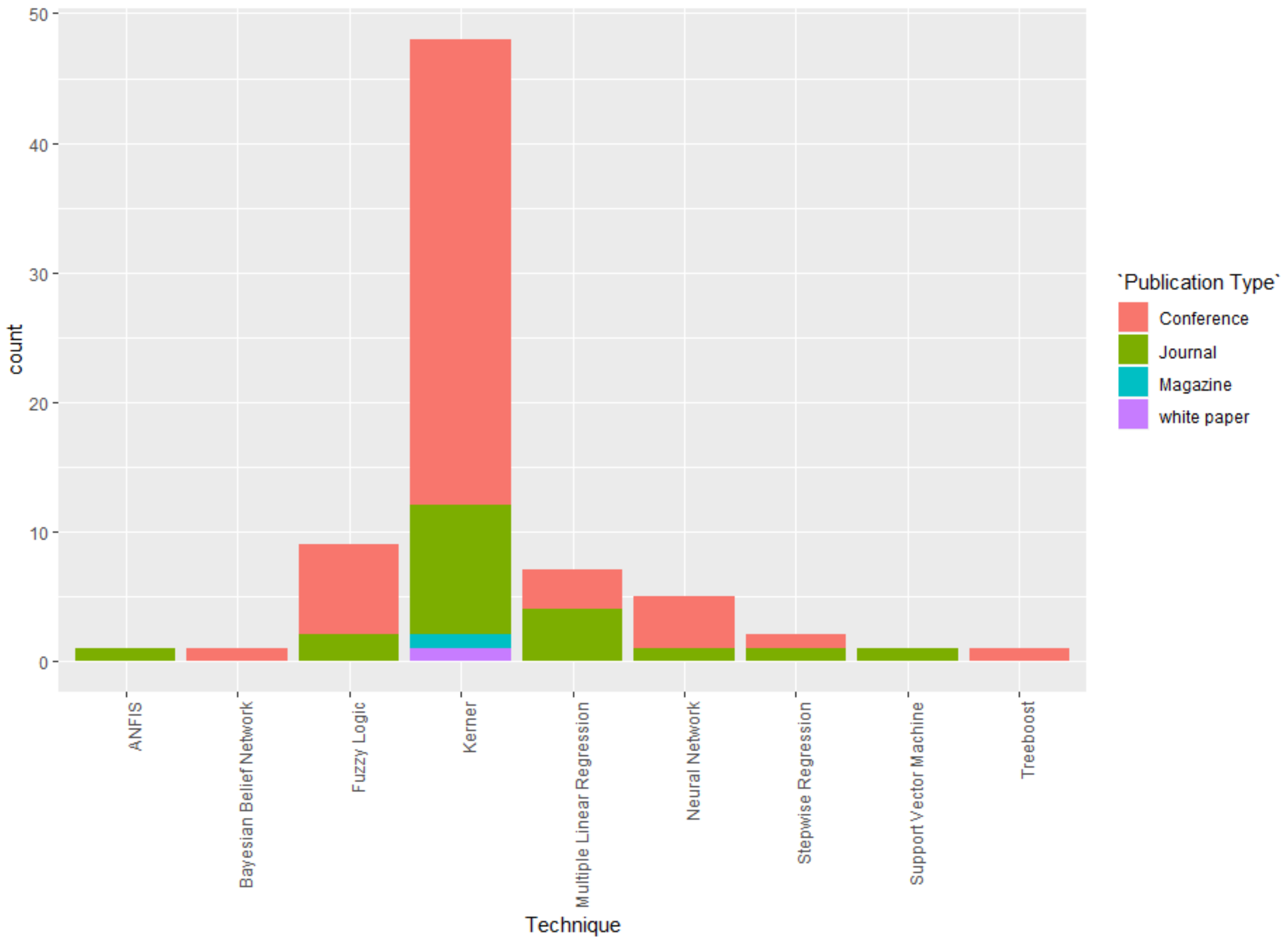

Fig8. Bar plot for main effort prediction techniques based on UCP.

4.5 MQ5. What are type of datasets that are used to run experiments for UCP effort estimation?

Until recently, most studies were using data collected from case studies or small number of business projects. This was clearly appearant in 90's and 2000's because most software companies do not document use case diagrams or sometimes, they do not follow a particular guideline when writing use case description. The case studies usually proposed by the authors of the papers based on information collected from some available project (either educational or commercial). Note that industrial datasets are those that collected within company from real business projects. While educational projects are those that collected by students. Some authors prefer collecting data from students because their graduation projects were most likely available and easy to collect after course of explanation and rehearsal.

From the analysis we found 62 (82.7%) of the selected papers used data in their experiments. The rest of studies they just propose ideas or discuss some issues related to UCP model. The list of all studies that used data to validate their hypothesis or models are listed in Table 9. The table illustrates the type of dataset that was used in the selected studies. From these studies we found that both industrial datasets and case studies were the dominant type of datasets, where 41.9% of the studies used case studies to validate the hypothesis, and 35.5.3% of the studies used industrial datasets. 19.4% of the papers used both industrial and educational studies. The most interesting observation is that most studies during the period 1993 until 2005 used few case studies or small dataset to validate their work. The first reason for that is the UCP method was not of greater research interest among researchers. The second thing is that the UCP itself was not widely spread among practitioners because it needs a lot of details regarding translation of use case diagram to metric values, in addition to assessment of environmental and technical variables that are subject to some degree of uncertainty.

Table 9. Studies classification based on type of datasets.

| "Dataset Type | Studies | Proportion |
|---|---|---|
| Industrial dataset | [S11][S14][S19][S23][S24][S26][S29][S32][S33][S34][S37][S40][S44][S46][S47][S52][S53][S57][S63][S64][S70][S72] | 35.5% |
| Educational Dataset | [S31][S34] | 3.2% |
| Both Industrial and Educational | [S35][S36][S38][S39][S41][S51][S54][S56][S62][S68][S69][S73] | 19.4% |
| Case Studies | [S1][S2][S4][S5][S6][S7][S8][S10][S12][S15][S16][S21][S27][S28][S30][S42][S43][S45][S58][S59][S60][S61][S65][S66][S71][S75] | 41.9%" |

Figure 9. shows Bar plot of the number of projects in each case study. It is clear that most of the case studies build their knowledge on data of one project which is empirically not acceptable to generalize their findings. This can be also confirmed in Table 10 which shows the studies and percentages for each used number of projects in their case studies.

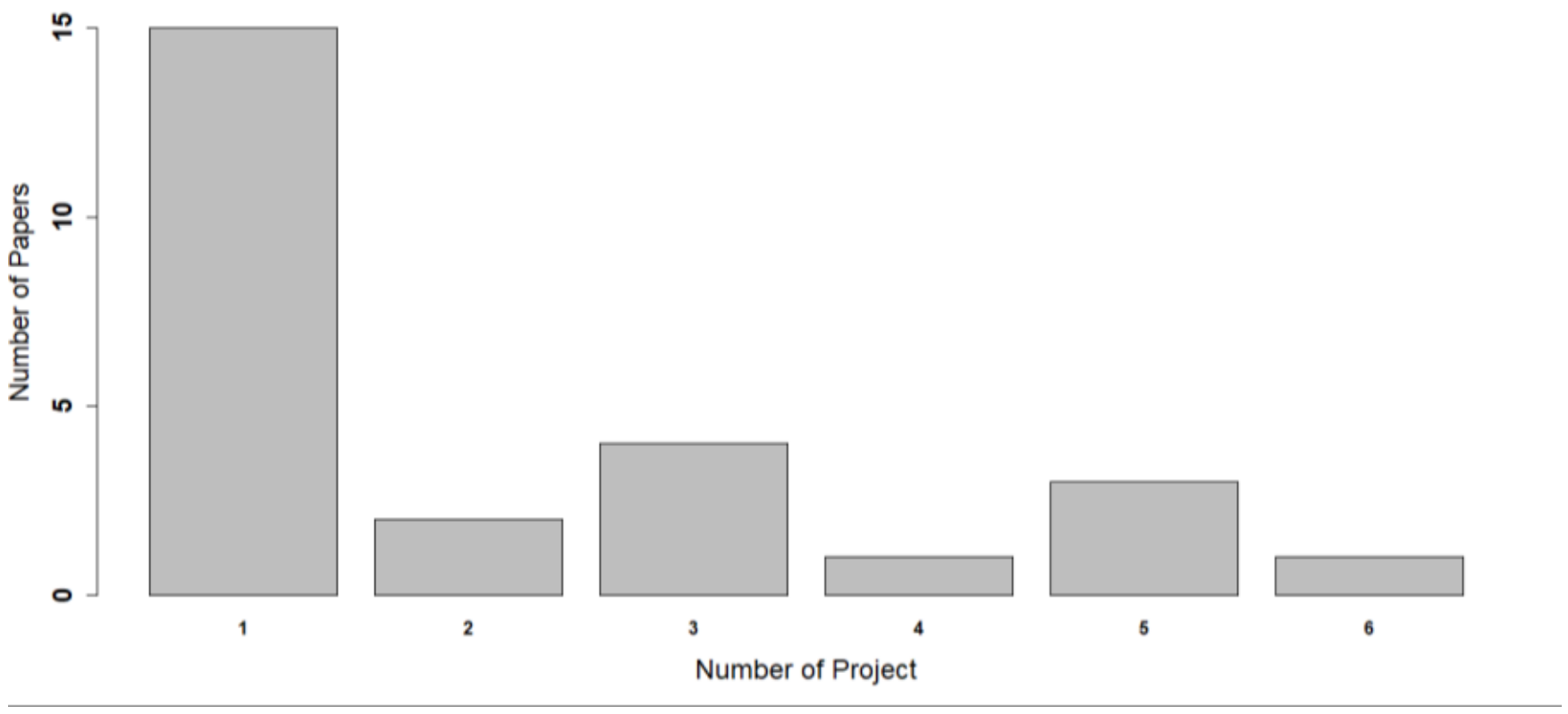

Fig. 9 Bar plot of number of case studies and number of its projects

Table 10. Case Studies classification based on number of projects.

| # Projects | Studies | Proportion |
|---|---|---|
| 1 Project | [S10][S16][S21][S28][S30][S42][S43][S45][S58][S59][S60][S65][S66][S71][S75] | 57.6% |
| 2 Projects | [S15] [S61] | 7.7% |
| 3 Projects | [S1] [S2][S4] [S8] | 15.4% |
| 4 Projects | [S27] | 3.8% |
| 5 Projects | [S5][S6] [S12] | 11.5% |
| 6 Projects | [S7] | 3.8% |

## 5. Results of Review Questions

This section presents and comments on the findings obtained from mapping questions. Specifically, we present answers to our proposed review question with the aim to synthesis the findings from our analysis in section 4.

### 5.1 RQ1. How easy is to collect UCP data from software industry and what is the implication on effort estimation?

The reported accuracies of the selected studies depend on two parameters: (1) quality of dataset including the number of employed projects; and (2) the used evaluation method (leave-one-out cross validation, holdout, n-fold cross validation, evaluation criteria, etc.). With respect to first point, we have seen from the results of MQ5, the evaluation of UCP effort estimation is primarily performed on case studies with a percentage of 41.9% or industrial UCP datasets with percentage of 35.5% of the selected papers. Also, most of case studies used a few numbers of projects to validate their hypothesis as mentioned in section 4.5. A few of them reported accuracies in terms of the well-known accuracy measures. Table 11 shows the most frequently datasets and their availability. It is important to note that multiple datasets were used in some studies. Interestingly, Nassif dataset is the most frequently used (63%), followed by the Ochodek dataset (45.5%). Note that the review takes into account both industrial and educational datasets. The most remarkable part of this analysis is that not all datasets are available for community to replicate experiments and generalize extracted knowledge. Unlike other effort estimation research areas, this is a big challenge that face this area of interest.

Table 11. Summary of main data sources that were used in previous studies

| Main Author | Studies | Prop% | # Projects | Source | Note | Availability |
|---|---|---|---|---|---|---|
| Ochodek | [S35][S44][S52][S53][S56][S62] [S63][S68][S69][S70] | 45.5% | 14 | [S35] | It contains 7 student projects and 7 industrial projects. | √ |
| Nassif | [S24][S26][S29][S36][S38][S39] [S41][S51][S54][S56][S57][S62][ S68][S69] [S73] | 63% | 110 | [S41] | It contains 65 student projects and 45 industrial projects. | × |
| Silhavy | [S63][S69][S70] | 13.6% | 73 | [S63] | | √ |
| Frohnhoff | [S14][S33] | 9.1% | 15 | [S14] | The 15 projects were developed by the IT-company. | ± |
| Robiolo | [S19] | 4.5% | 13 | [S19] | | ± |
| Lavazza | [S23] | 4.5% | 17 | [S23] | 17 project were developed in three different environments. | ± |
| ARUMUGAM | [S34] | 4.5% | 15 | [S34] | 15 object oriented projects developed in SE laboratory | √ |
| ALWIDIAN | [S37] | 4.5% | 14 | [S37] | 14 Industrial projects | √ |
| Iraji | [S40] | 4.5% | 5600 | [S40] | Derived from ISBSG projects | × |
| Alves | [S46] | 4.5% | 7 | [S46] | | ± |
| Badri | [S64] | 4.5% | 5 | [S64] | 5 open java source project | × |

√: public dataset and available for community, ×: private dataset, and may available on request. ±: partial available

Regarding evaluation techniques that have been followed in the selected studies, not all selected studies used the common machine learning validation techniques (e.g. Leave one-out cross validation, Hold-out and K-Fold

Cross validation) because they mainly use a few number of projects as case studies and they did not apply either machine learning or data mining techniques. However, the studies that used validation strategies are listed in Table 12 and categorized into three types of validations. The most popular of these were Hold-out and K-fold cross validation (k > 1) with percentage of 35.7% each.

Table 12. Summary of validation techniques.

| Validation Technique | Studies | Proportion |
|---|---|---|
| Leave-One-Out Cross Validation (LOOCV) | [S64][S66][S70][S71] | 28.6% |
| Hold-out | [S26][S28][S31][S34][S72] | 35.7% |
| k-Fold Cross Validation | [S37][S40][S41][S43][65] | 35.7% |

Regarding the accuracy measures that were used to validate UCP effort estimation models and their variants, there was a discrepancy in the type of evaluation measures. Table 13 describes the most common accuracy measures the are frequently used to validate UCP effort estimation models. The most noticeable observation is that only 44% of the papers recorded accuracies in terms of the well-known accuracy measures as shown in Table 14. Particularly, MAE was used in 9 of the studies (12%), MMRE was used in 19 of the studies (25.3%), MMER was used in 4 of the studies (5.3%), Pred(25) was used in 13 of the studies (17.3%), MBRE and MIBRE were used in 6 of the studies each (8%) and MSE was used in 7 of the studies (9.3%).

Table 13. Description of the frequent evaluation measures.

| "Name | Equation | Description |
|---|---|---|
| Mean Absolute Error | $MAE = \frac{\sum_{i=1}^{n}\lvert e_i - \hat{e}_i\rvert}{n}$ | To measure the average of errors. |
| Mean Magnitude Relative Error | $MMRE = \frac{1}{n}\sum_{i=1}^{n}\frac{\lvert e_i - \hat{e}_i\rvert}{e_i}$ | To measure the average of relative errors to actual projects effort. |
| Magnitude of relative error relative to the estimate | $MMER = \frac{1}{n}\sum_{i=1}^{n}\frac{\lvert e_i - \hat{e}_i\rvert}{\hat{e}_i}$ | To measure the average of relative errors to predicted projects effort. |
| Mean Balanced Relative Error | $MBRE = \frac{1}{n}\sum_{i=1}^{n}\frac{\lvert e_i - \hat{e}_i\rvert}{min\ (e_i, \hat{e}_i)}$ | To measure the average of relative errors to minimum between actual and predicted effort. |
| Mean Inverse Balanced Relative Error (MIBRE) | $MIBRE = \frac{1}{n}\sum_{i=1}^{n}\frac{\lvert e_i - \hat{e}_i\rvert}{max\ (e_i, \hat{e}_i)}$ | To measure the average of relative errors to maximum between actual and predicted effort. |
| Standardized Accuracy | $SA = 1 - \frac{MAE}{\overline{MAE}_{po}}$ | To test whether the prediction model really surpasses a baseline of random guessing and produces meaningful predictions. Where $\overline{MAE}_{po}$ the mean absolute errors of random guessing. |
| Performance | $PRED(\ell) = \frac{\lambda}{n} \times 100$ | To count the percentage of relative errors that fall within less than or equal to$\ell$ of the actual values. where $\lambda$ is the count of project that have magnitude relative error less than $\ell$. |
| Root Mean Squared Error | $RMSE = \sqrt{\lvert e_i - \hat{e}_i\rvert^2}$ | To measure Mean Squared Error." |

Where $e_i$ and $\hat{e}_i$ are the actual and predicted effort respectively.

Table 14 summarizes the frequent accuracy measures for all studies that recorded their evaluation results. We have noticed that not all studies share the same accuracy measures during their validation, taking in account that these studies use different versions of the dataset. For studies with different model configurations we used optimal configuration if possible, otherwise we use the average of accuracy values if there were different dataset samplings. To further analyze the distribution of the accuracy measures of UCP effort estimation techniques, we drew interval plots corresponding to each of these criteria using the estimation accuracy values of each selected study. As can be seen in Figure 10, the mean of the accuracy values of UCP effort estimation techniques are around 31.5% for MMRE, 32.1% for MMER, 80.6% for Pred (25), 27.0 for MBRE, 15.7 for MIBRE, 102.7 for RMSE, 2093 for MAE and 72.7% for SA. However, it can be noticed that, according to the distribution analysis of MMRE, MBRE, MIBRE and RMSE criteria, the UCP effort estimation techniques are positively skewed, While the distribution of MMER and SA are symmetrically distributed around the median. Whereas the Pred is negatively skewed. In addition, there is high variability in the distribution of Pred and MMER. The results obtained by these studies indicate that UCP methods tend to produce acceptable estimates, especially if we consider this type of estimate is done at early stage of software development.

Table 14. Summary of reported accuracy measures.

| # Study ID | MMRE | MMER | PRED | MBRE | MIBRE | RMSE | MAE | SA | Model | # instances | Data Type |
|---|---|---|---|---|---|---|---|---|---|---|---|
| S3 | 21% | - | | - | - | - | - | - | Algorithmic | 1 | CS |
| S15 | 30.5% | - | 62.5% | - | - | - | - | - | Algorithmic | 2 | CS |
| S19 | - | - | - | - | - | 129.92 | - | - | Algorithmic | 13 | Ind |
| S20 | - | - | - | - | - | - | 4.7 | - | ANN | 75 | Ind |
| S21 | - | - | - | - | - | 329 | - | - | Algorithmic | 1 | CS |
| S23 | 25.6% | - | 55.2% | - | - | - | - | - | Algorithmic | 17 | Ind |
| S24 | 45% | 27.6% | - | - | - | - | - | - | FL + ANN | 20 | Ind |
| S26 | 24% | - | 95.8% | - | - | - | - | - | FL | 24 | Ind |
| S29 | 23% | - | 95.8% | - | - | - | - | - | FL | 24 | Ind |
| S35 | 40% | - | 50% | - | - | - | - | - | Regression | 14 | Ind+ Edu |
| S36 | 29% | - | 64% | - | - | - | - | - | Treeboost + Regression | 59 | Ind+ Edu |
| S38 | 49% | - | 86.11% | - | - | - | - | - | ANN | 240 | Ind+ Edu |
| S39 | - | 54% | 89% | - | - | - | - | - | ANN | 72 | Ind+ Edu |
| S40 | - | 7% | - | - | - | - | - | - | FL | 7 | Ind |
| S41 | - | 40% | 92% | - | - | 24 | 8940 | | Regression | 110 | Ind+ Edu |
| S44 | 9% | - | - | - | - | - | - | - | Algorithmic | 14 | Ind |
| S46 | 66.3% | - | - | - | - | - | - | - | Algorithmic | 7 | Ind |
| S51 | 38.6% | | 98% | | | | | | SVM | 84 | Both |
| S52 | 44.5% | - | - | - | - | - | - | - | FL | 14 | Ind+ Edu |
| S54 | - | - | - | 15.9% | 12.5% | - | 2231.8 | - | FL | 14 | Ind |
| S56 | - | - | - | 17.5% | 13.2% | - | 1598 | 84.3% | Algorithmic | 195 | Ind+ Edu |
| S57 | 34% | - | 97.3% | - | - | - | - | - | Algorithmic | 149 | Ind+ Edu |
| S61 | 6.5% | - | - | - | - | - | - | - | Algorithmic | 10 | Ind |
| S62 | - | - | - | 16.8% | 13.1% | - | 3182 | 62% | Algorithmic | 2 | CS |
| S63 | - | - | - | - | - | 18.74 | - | - | Algorithmic | 98 | Ind+ Edu |
| S64 | - | - | - | - | - | 58.71 | 33.7 | - | Stepwise multiple linear regression | 5 | Ind |
| S68 | - | - | - | 65.9% | 24.1% | - | 1761.4 | 73.15% | simple linear regression | 195 | Ind+ Edu |
| S69 | - | - | - | 28% | 18% | - | 820.4 | 71.3% | Algorithmic | 195 | Ind+ Edu |
| S70 | - | - | - | - | - | 55.9 | - | - | Algorithmic | 98 | Ind+ Edu |
| S71 | 18.4% | - | - | - | - | - | - | - | Algorithmic | 1 | CS |
| S72 | 30.4% | - | 82% | - | - | - | - | - | Regression | 22 | Ind+ Edu |
| S73 | - | - | - | 17.7% | 13% | - | 262.8 | - | Algorithmic | 110 | Ind+ Edu" |

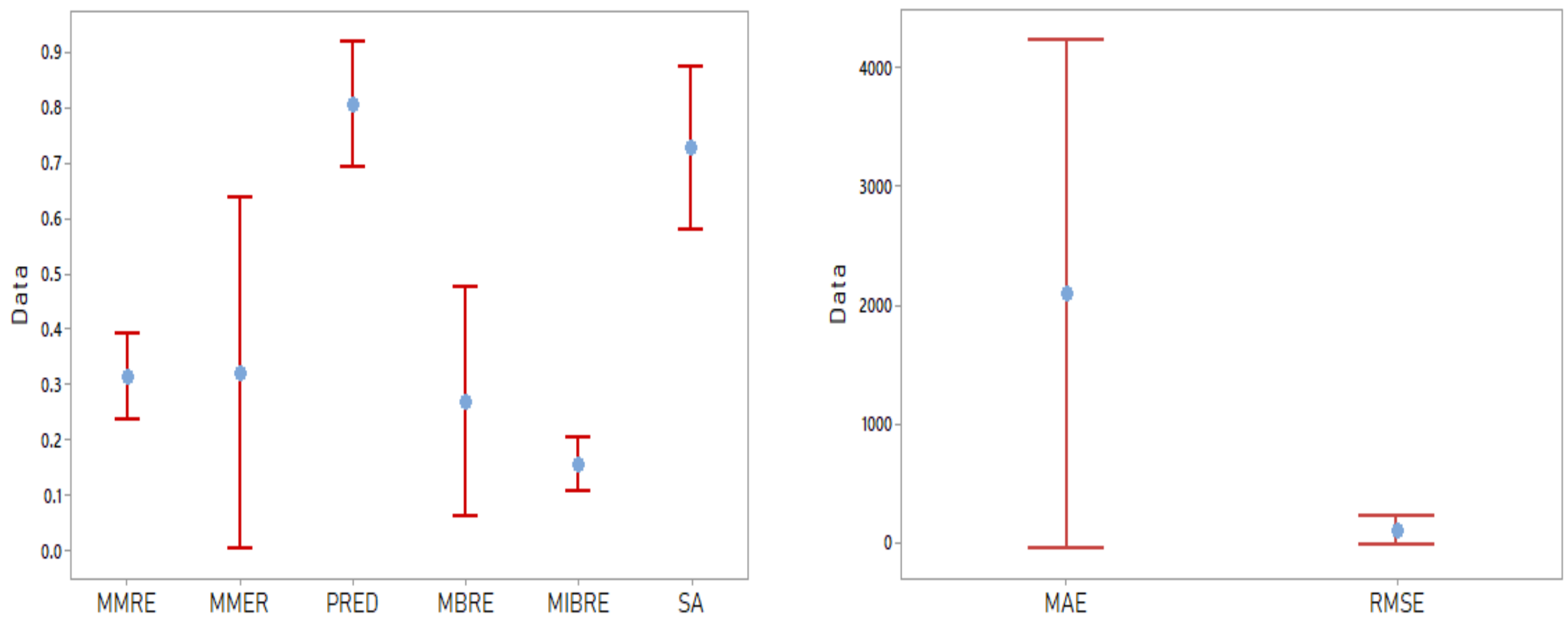

Figure 10. Interval Plot of accuracy measures for all studies in Table 12

To further analyze the estimation accuracy of UCP methods, Table 15 provides the detailed statistics of the most accuracy measures MMRE, Pred(25), MBRE and MIBRE for each of the most frequently used datasets (Nassif and Ochodek). Corresponding to Nassif dataset, the mean of the prediction accuracy values varies from 23% to 49% for MMRE, from 64% to 98% for Pred(25), from 15.9% to 65.9% for MBRE and from 12.5 to 24.1 for MIBRE. For Ochodek dataset, the mean of the prediction accuracy values varies from 9% to 44.5% for MMRE, from 16.8% to 65.9% for MBRE and from 13.1 to 24.1 for MIBRE. This indicates that UCP methods tend to yield acceptable estimates.

Table 15. Summary of reported accuracy measures for top two frequent datasets.

| "Dataset | MMRE | | | | Pred | | | | MBRE | | | | MIBRE | | | |
|---|---|---|---|---|---|---|---|---|---|---|---|---|---|---|---|---|
| | Min% | Mean% | Median% | Max% | Min% | Mean% | Median% | Max% | Min% | Mean% | Median% | Max% | Min% | Mean% | Median% | Max% |
| Nassif | 23% | 35% | 34% | 49% | 64% | 90% | 94% | 98% | 16% | 27% | 18% | 66% | 13% | 16% | 13% | 24% |
| Ochodek | 9% | 31% | 40% | 45% | 50% | 50% | 50% | 50% | 17% | 32% | 23% | 66% | 13% | 17% | 16% | 24%" |

## 5.2 RQ2. What are the impacts of combining other techniques with an UCP method technique on its estimation accuracy?

Since the introduction of the original UCP model in 1993 by Kerner, various of machine learning and data mining technique have been adopted and combined with UCP in order to improve its accuracy. The uses of such techniques were utilized for two goals: the first one is to improve the UCP sizing procedure, while the second goal is to build effort estimation model based on UCP variable. However, since the use of such technique enhance UCP sizing approach, it will have direct impact on the accuracy of UCP effort estimation. From Table 14 we summarize the results of accuracies for all techniques that were used either to enhance UCP size measure or to improve accuracy of effort estimation. The studies that followed Kerner approach in measuring UCP and estimating effort are called algorithmic in Table 16. From the table, we can notice that all used techniques produce relatively acceptable accuracy. If we consider the median statistic method as reference for comparison because it is less bias than other statistical measures, we notice that algorithmic model is the superior method in terms of MMRE and RMSE, whereas Fuzzy Logic is the superior models in terms of Pred, MBRE and MIBRE. ANN is best technique in terms of MAE. Note that each model may use different dataset and different validation technique. However, this can confirm that there is instability ranking among these techniques. Also, the use of machine learning methods such as Neural networks and Regression do not always produce superior predictions in comparison with Algorithmic techniques. This can open a new research direction to assess the usefulness of machine learning for predicting UCP effort estimation especially when historical datasets are become available.

The benefits of combining UCP with other techniques, especially machine learning, can still be useful for improving productivity prediction as it is an important key driver in translating UCP into the most likely software effort. In addition, measuring use case complexity can also benefit from combination with other techniques such as Neural Networks and Fuzzy Logic.

Table 16. Summary of accuracy measures for the most used estimation techniques

| Accuracy Measure | Statistic method | Algorithmic | Fuzzy Logic | Neural Networks | Regression |
|---|---|---|---|---|---|
| MMRE% | Min | 7.0 | 23.0 | 49.0 | 30.0 |
| | Mean | 26.5 | 34.25 | 49.0 | 35.0 |
| | Median | 23.5 | 34.5 | 49.0 | 35.0 |
| | Max | 66.0 | 45.0 | 49.0 | 40.0 |
| Pred% | Min | 55.0 | 96.0 | 86.0 | 50.0 |
| | Mean | 71.7 | 96.0 | 875 | 74.7 |
| | Median | 63.0 | 96.0 | 87.5 | 82.0 |
| | Max | 97.0 | 96.0 | 89.0 | 92.0 |
| MBRE% | Min | 17.0 | 16.0 | - | 66.0 |
| | Mean | 20.3 | 16.0 | - | 66.0 |
| | Median | 18.0 | 16.0 | - | 66.0 |
| | Max | 28.0 | 16.0 | - | 66.0 |
| MIBRE% | Min | 13.0 | 13.0 | - | 24.0 |
| | Mean | 14.3 | 13.0 | - | 24.0 |
| | Median | 13.0 | 13.0 | - | 24.0 |
| | Max | 18.0 | 13.0 | - | 24.0 |
| MAE | Min | 262.8 | 2231.8 | 4.7 | 33.7 |
| | Mean | 1465.8 | 2231.8 | 4.7 | 3578.4 |
| | Median | 1209.2 | 2231.8 | 4.7 | 1761.4 |
| | Max | 3182 | 2231.8 | 4.7 | 8940 |
| RMSE | Min | 18.74 | - | - | 24.0 |
| | Mean | 133.4 | - | - | 41.4 |
| | Median | 92.9 | - | - | 41.4 |
| | Max | 329 | - | - | 58.7 |
| SA% | Min | 62.0 | - | - | 73.0 |
| | Mean | 72.3 | - | - | 73.0 |
| | Median | 71.0 | - | - | 73.0 |
| | Max | 84..0 | - | - | 73.0 |

### 5.3 RQ3. What is the most favorable context for UCP effort estimation method?

Authors in the area of UCP effort estimation work on different contexts to improve performance and reliability of UCP method. Therefore, it is important to identify the favorable contexts to direct researchers on the most interesting research parts of such method. Ochodek et al. [1] and Nassif et al. [3] identified several factors that affect the UCP method such as complexity measure of use cases, assessment of UCP adjustment factors, impact of learning productivity. Based on their findings, we attempt to identify the favorable UCP contexts that received greater attention than others. To accomplish that we extracted and investigated the strengths and weaknesses reported in the selected UCP studies as shown in Tables 15 and 16.

We found that the information reported is mainly related to UCP sizing technique and productivity factor prediction, which seem to have a significant impact on the prediction accuracy of UCP method. Although the philosophy behind construction of UCP method was inspired by Function Points method, the construction of UCP was broadly discussed and questioned because of the algebraic construction of all metrics and the weights that are used in the calculation. For instance, researchers turn attention to the limitation on the construction of UCP such as technical complexity factors assessment, environment factors assessment and involvement of calculations that are based on algebraically inadmissible scale-type transformations. Yet, there is no study attempts to validate the weights that are used in computing UAW, UUCW, EF and TCF metrics [21] and [22]. Authors of [23], [24], [25], [26], [4] and [1] went further to extend the UCP model by providing new complexity weights or by modifying the method used to predict effort. Other researchers work on the experience needed to translate use case descriptions into suitable metric (i.e. UUCW). Few authors discusses different approaches to measure the complexity of use cases based on transaction, paths and TTPoints [1], [23]–[25], [31] and [31]. Transaction is a set of trigger-response activities that is extracted from textual description of use cases. Paths is another measure based on cyclomatic complexity metric which is also derived from textual description of the use cases. In this regard, Robiolo et al., [27][29] proposed an improvement to transactions by calculating paths which are computed form the cyclomatic complexity of the use case scenario. They introduced the concept of stimuli which is a system entry point that generates response (transaction) of an actor action in a use case. Ochodek et al. [1] discusses a reliability of transaction identification process in line of other use case complexity measures such as TTPoints. The main conclusion drawn from these studies is that both paths and transaction are useful for UCP calculation even that there is a slight difference in the accuracy. However, there is no automated tool that can help in this issue.

Recently, authors of [3], [19], [20], [32] studies shade the light into the need to learn the productivity factor from environmental factors of historical projects. They discussed the importance of using dynamic productivity ration instead of using fixed productivity ratio that was suggested by Kerner. Other approaches suggest learning productivity from UCP size metrics which could be as alternative solutions to improve accuracy of translating UCP into most likely software effort.

There are characteristics other than UCP construction parameters to be considered when using a UCP method. We summarize these in Tables 17 and 18. For example, an UCP technique is well defined procedure because it depends on robust modelling technique which is UML. This provided UCP the support needed to be spread across industry because use case diagram is a de-facto analysis technique for object-oriented software projects. In addition, the use case specification always receives support from software engineering community which helps to keep UCP model more stable. Moreover, the UCP has proven to be more accurate than expert judgement especially at early stages of software development. This is very intuitive as the UCP depends heavily on the use case diagram that has been extracted from software requirements. On the other hand, UCP is still not widely used within agile project development because of the nature of agile projects that depend on dividing project into small tasks with small number of requirements to be built within a period called sprint. This approach hinders applying use case diagram on a very small set of the project requirements while ignoring other important requirements.

To summarize the findings, the UCP is promising approach for early effort estimation and has many benefits for software industry. But still some improvements are required in some sides in order to cover all issues mentioned in Table 18. As mentioned in the previous section combining UCP with machine learning techniques could be a solution, for some limitations mentioned in Table 18, when we carefully consider the characteristics of the UCP dataset.

Table 17. List of main advantages of applying UCP with supporting studies

| “Advantage | Supporting studies |
|---|---|
| 1-Improving early effort estimation | [S4][S7][S8][S9][S10][S11][S22] [S23][S24][S26][S29][S33][SS39][S40] [S42][S46][S51][S56][S57][S58][S60] [S62][S63][S64][S68][S69][S70][S73] |
| 2-UCP is well defined procedure | [S31][S33][S48][S56][S57][S62] |
| 3-UCP performs better than algorithmic models at early stage of software development | [S38][S44][S61] |
| 4-UCP performs better than expert judgement | [S2][S3][42][S61] |
| 5-UCP effort estimation method does not always need historical data to produce effort estimate | [S13][S16][S43][S47][S62][S68][S69]” |

Table 18. List of main limitations mentioned in the published UCP with supporting studies

| "Limitation | Supporting studies |
|---|---|
| 1-There is no standard for use case specification | [S27][S33][S43][S60] |
| 2-Assessment of Technical Factors and Environmental Factors are subject to expert some degree of uncertainty | [S35][S61][S62][S68][S69] |
| 3-Algebraic constructions of UCP methods and arbitrary factors weight are not clear. | [S35] |
| 4-There is not stable technique to translate UCP into corresponding effort | [S29][S33][S36][S38][S39] [S41][S56][S62] |
| 5-Measuring Use Case complexity using transaction is not always accurate in comparison to paths and TTPoints | [S19][S23][S35] |
| 6- UCP cannot be easily adopted in Agile Development | [S10][S42][S45][S58]" |

## 5.4 RQ4. To what extent are researchers on UCP methods aware of the breadth of potential estimation study sources.

Our main goal in this question is to make sure that researchers are aware of the variety of possible venues to publish studies related to the UCP method. An indication of this awareness was derived through a random selection of 5 UCP effort estimation journal papers and 10 conference papers (about 20% of the total). These 15 papers are marked with (A) in Appendix A. We noticed that: 1) Researchers usually do not refer to previously published, papers on the same research topic. This can be clearly seen in their related work or comparison with previous published results. Some researchers are not aware with state-of art results or models published with area of interest. 2) Conference papers are cited more than journal papers. This explain why the contributions of the new studies do not really handle the major challenges in this area of interest. For example, we can see too many repetitions in the experiment without drawing useful conclusion of clear contribution. We found that 40 percent of papers made references to at least one of the top five journals mentioned in Table 5. There were, relative to the number of papers available, few references to papers in IET software (20 percent), Innovations in Systems and Software Engineering, or Journal of Software: Evolution and Process. Most of citation were made to both Elsevier Journals: Information and Software Technology and Journal of Systems and Software. 3) Few papers referred to estimation results outside the UCP research community, e.g., to studies in soft computing, and artificial intelligent. The main portion of references to sources outside the software community seems to be to literature on fuzzy logic.

# 6. Discussion and Research Implication

This section presents summary of results and recommendations for future research. With respect to main type of contributions in the field of UCP effort estimation, our review revealed that the enhancement and validation on UCP model are the most frequent contribution types. Enhancement aims to improve the structure of UCP model to produce more accurate estimates, while validations aims at evaluating of the performance of existing UCP techniques using some historical datasets. Our review revealed that no one of the enhanced UCP models was adopted later as de-facto model for further investigation. This raise a concern regarding the quality and performance of such models. Therefore, it is recommended that future studies replicate the previous enhanced UCP models on various software development contexts and using new industrial datasets. Moreover, the review has found that there is a lack of in-depth studies on how to evaluate UCP models in real-life contexts.

In addition, most of the cases studies and datasets that were used to evaluate the UCP model are too obsolete to be representative. We suggest that researchers should put more effort on collecting reliable and quality UCP datasets which can help in evaluating and enhancing UCP models. In this regard, the UCP method can benefit from other related research fields that aims to improve UML modelling and specification. In fact, the use case specification has great impact on how the use case complexity are calculated which therefore affect final estimate accuracy. On the other hand, few tools implementing UCP method have been developed. It is perhaps not surprising that the use of UCP method within software industry is so limited. To solve this issue, we recommend that researchers implement automated tools to translate use case diagram and its specification to size metrics and provide guidelines on how to use these tools in industry.

Regarding the main techniques that have been used to produce effort estimates from UCP. This review revealed that fuzzy logic, multiple linear regression, and neural network are the frequent employed technique. Researchers put further focus on fuzzy logic because of imprecision nature of data collected at early stage of software development. However, we recommend that other techniques can be used in combination of UCP to enhance its accuracy, for example optimization algorithms can be used to search for best environmental and technical factors that improve UCP size metric. Some other techniques, such as genetic programming and Bayesian networks were not used in combination with UCP method. Therefore, researchers are encouraged to investigate the impact of these techniques when combined with UCP.

With respect the estimation accuracy, this review revealed that most UCP models tend to produce acceptable accuracy level. However, these results are based on case studies with few examples or on confidential datasets. Therefore, it is recommended to run more experiments on large industrial datasets. Also, there were no common accuracy metrics among researchers. Researchers are encouraged to use the common accuracy measures that proved to avoid bias and can generalize findings in the future such as Standardized accuracy (SA), MdMRE, MBRE and MIBRE. Researchers are also encouraged to discard biased measures such as MMRE and MMER.

Regarding datasets, our review showed that the cases studies and small datasets were the most common source of UCP evaluation. This shortage in reliable datasets, especially industrial one (i.e. which collected within software companies) hinder the generalization of UCP among practitioners and researchers. We recommend that researcher should work in line with software industry to build a comprehensive data collection tools that attempt to analyze use case diagram and translate its specification into appropriate UCP metrics values.
Regarding the estimation contexts of UCP method, researchers should be aware of the impact various technical and environmental factors assessment on the constructing and evaluating UCP models. In addition, researchers should be aware of the best way to compute the complexity of use cases. Although the common technique to measure the use case complexity is based on transactions, our review revealed that there are different versions of measuring use case complexity such as paths and TTPoints that proven to produce better results than transactions in some studies. In this regard, we suggest that researchers should put more effort on improving the construction of UCP such as technical complexity factors assessment, environment factors assessment and involvement of calculations that are based on algebraically inadmissible scale-type transformations. few research works have studied the limitations of applying UCP in agile software development. It would be beneficial for the research community to address this limitation since the existing structure of UCP method cannot be easily adopted with agile development.

Finally, the set of studies are dominated by 12 researchers who have appeared in at least three studies each as shown in Table 19. In particular, we found three main research groups which contributed good number of publications. The first group involve researchers from University of Western Ontario in Canada which has been involved in 10 of the studies: Ali Bou Nassif (16), Luiz Capretz (10) and Danny Ho (10). These papers are published in different venues, mainly on IEEE. Mohammad Azzeh and Ali Bou Nassif worked together in 8 studies (4 conference papers and 4 journal papers published in Science Direct, IET and Wiley). Bente Anda contributed with 5 studies in this research field. Petr Silhavy and Radek Silhavy were also another team who contributed with three publications, mainly on Science Direct. Furthermore, Miroslaw Ochodek at Poznan University of Technology has a substantial contribution in software requirements that are related to the applications of use cases in general. However, he has limited studies that address the UCP method in particular.

Table 19. Summary of accuracy measures for the most used estimation techniques, where C: Conference, J: Journal

| Author name | ACM | | Science Direct | | Google Scholar | | IEEE | | IET | Springer | | Wiley | Total |
|---|---|---|---|---|---|---|---|---|---|---|---|---|---|
| | C | J | C | J | C | J | C | J | J | C | J | J | |
| Ali Bou Nassif | 1 | 0 | 0 | 2 | 2 | 1 | 7 | 0 | 1 | 0 | 0 | 2 | 16 |
| Bente Anda | 1 | 0 | 0 | 0 | 0 | 0 | 1 | 0 | 0 | 3 | 0 | 0 | 5 |
| Danny Ho | 1 | 0 | 0 | 1 | 2 | 1 | 5 | 0 | 0 | 0 | 0 | 0 | 10 |
| Ford Gaol | 0 | 0 | 0 | 0 | 0 | 0 | 3 | 0 | 0 | 0 | 0 | 0 | 3 |
| Gabriela Robiolo | 2 | 0 | 0 | 0 | 0 | 0 | 0 | 0 | 0 | 0 | 1 | 0 | 3 |
| Harco Warnars | 0 | 0 | 0 | 0 | 0 | 0 | 3 | 0 | 0 | 0 | 0 | 0 | 3 |
| Jerzy Nawrocki | 0 | 0 | 0 | 2 | 0 | 1 | 0 | 0 | 0 | 0 | 0 | 0 | 3 |
| Luiz Capretz | 1 | 0 | 0 | 1 | 2 | 1 | 5 | 0 | 0 | 0 | 0 | 0 | 10 |
| Mohammad Azzeh | 0 | 0 | 0 | 1 | 0 | 0 | 4 | 0 | 1 | 0 | 0 | 2 | 8 |
| Petr Silhavy | 0 | 0 | 0 | 2 | 0 | 0 | 0 | 0 | 0 | 1 | 0 | 0 | 3 |
| Radek Silhavy | 0 | 0 | 0 | 2 | 0 | 0 | 0 | 0 | 0 | 1 | 0 | 0 | 3 |
| Shadi Banitaan | 0 | 0 | 0 | 0 | 0 | 0 | 2 | 0 | 1 | 0 | 0 | 0 | 3 |

## 7. Conclusions

Predicting software effort from UCP size metric has witnessed increase interest among researchers in the past two decades. This kind of estimation is designed mainly for early software estimation. Although this approach has promises for better early effort estimation, there are many challenges that hinder the spread of this method in the industry. This systematic literature reviews most relevant published studies in this area of interest with attempts to focus the light on the main challenges and potential research directions. Among 127 searched papers, only 75 relevant papers with good quality have been selected for investigation. Twenty out of 75 selected papers were published in journals which depicts that the quality of work published in the field of interest is still low. The main reasons are that: 1) absence of reliable public industrial datasets, 2) most authors mainly focus on building effort estimation from UCP, but few studies were designed to improve process of UCP, 3) The algebraic construction of UCP method has not been toughly validated, mainly the arbitrary numbers used to adjust technical and environmental factors, and finally 4) most developers do not follow a proper guide when translating use case diagram into size metrics, which is easily influenced by uncertainty in understanding the use cases description. Another interesting observation is that half of the journal articles were published in nonspecialized software engineering journals. This may affect the visibility and quality of research outcomes of these studies. Even that the readers of these papers might not find them useful since they were affected with the theme of the journal that they published in.

Most articles used small numbers of projects because of the difficulty in obtaining reliable datasets that are commonly used in software cost estimation domains. This is raised from the fact that UCP data are different in structure from other effort estimation dataset. Furthermore, the process of measure the required information takes long time comparing with other effort estimation datasets. The good thing that we observed in this SLR is that large number of studies used industrial projects rather than case studies and educational projects.

The original UCP effort estimation that was proposed by Karner was the main the commonly used model in estimating effort from UCP, especially in the first decade after publishing UCP. This is because of the absent of historical datasets. Later, fuzzy logic and linear regression were among published effort estimation models.

Finally, Enhancement and building techniques were the dominant approaches for the UCP studies. There was little interest in developing estimation tools approach, especially intelligent tools that translate use case description to UCP metric. Also, we have investigated some journals to examine awareness of published results and breadth of potential estimation study sources. We found that authors of some journal articles, are not aware of previous published results and conclusions in the field of UCP effort estimation. Furthermore, Researchers usually do not refer to previously published, seemingly relevant, papers on the same research topic. Interestingly, few papers referred to estimation results outside the software community.

## ACKNOWLEDGMENTS

"Mohammad Azzeh is grateful to the Applied Science Private University, Amman, Jordan, for the financial support granted to cover the publication fee of this research article.

Ali Bou Nassif and Imtinan Attili would like to thank the University of Sharjah for supporting this research."

## References

[1] M. Ochodek, J. Nawrocki, and K. Kwarciak, "Simplifying effort estimation based on Use Case Points," *Inf. Softw. Technol.*, vol. 53, no. 3, pp. 200–213, 2011.

[2] M. Azzeh and A. B. Nassif, "A hybrid model for estimating software project effort from Use Case Points," *Appl. Soft Comput. J.*, vol. 49, 2016.

[3] A. B. Nassif, D. Ho, and L. F. Capretz, "Towards an early software estimation using log-linear regression and a multilayer perceptron model," *J. Syst. Softw.*, vol. 86, no. 1, 2013.

[4] P. Mohagheghi, B. Anda, and R. Conradi, "Effort estimation of use cases for incremental large-scale software development," in *Proceedings of the 27th international conference on Software engineering*, 2005, vol. St. Louis, pp. 303–311.

[5] A. B. Nassif, L. F. Capretz, D. Ho, and M. Azzeh, "A treeboost model for software effort estimation based on use case points," in *Proceedings - 2012 11th International Conference on Machine Learning and Applications, ICMLA 2012*, 2012, vol. 2, pp. 314–319.

[6] B. Kitchenham, O. Pearl Brereton, D. Budgen, M. Turner, J. Bailey, and S. Linkman, "Systematic literature reviews in software engineering - A systematic literature review," *Inf. Softw. Technol.*, vol. 51, no. 1, 2009.

[7] A. Idri, F. A. Amazal, and A. Abran, "Analogy-based software development effort estimation: A systematic mapping and review," *Information and Software Technology*, vol. 58. 2015.

[8] J. Wen, S. Li, Z. Lin, Y. Hu, and C. Huang, "Systematic literature review of machine learning based software development effort estimation models," *Inf. Softw. Technol.*, vol. 54, no. 1, pp. 41–59, 2012.

[9] M. Jorgensen and M. Shepperd, "A systematic review of software development cost estimation studies," *IEEE Trans. Softw. Eng.*, vol. 33, no. 1, 2007.

[10] G. Karner, "Resource Estimation for Objectory Projects," 1993.

[11] B. Kitchenham and S. Charters, "Guidelines for performing Systematic Literature Reviews in Software Engineering." 2007.

[12] Y. Mahmood, N. Kama, and A. Azmi, "A systematic review of studies on use case points and expert-based estimation of software development effort," *J. Softw. Evol. Process*, vol. pre-print, Jan. 2020.

[13] M. Usman, E. Mendes, F. Weidt, and R. Britto, "Effort estimation in Agile Software Development: A systematic literature review," in *ACM International Conference Proceeding Series*, 2014.

[14] S. MacDonell, M. Shepperd, B. Kitchenham, and E. Mendes, "How reliable are systematic reviews in empirical software engineering?," *IEEE Trans. Softw. Eng.*, vol. 36, no. 5, 2010.

[15] A. B. Nassif, L. F. Capretz, and D. Ho, "Enhancing Use Case Points Estimation Method using Soft Computing Techniques," *J. Glob. Res. Comput. Sci.*, vol. 1, no. 4, pp. 12–21, 2010.

[16] M. S. Iraji and H. Motameni, "Object Oriented Software Effort Estimate with Adaptive Neuro Fuzzy use Case Size Point (ANFUSP)," *Int. J. Intell. Syst. Appl.*, vol. 4, no. 6, pp. 14–24, 2012.

[17] M. Azzeh, A. B. Nassif, and S. Banitaan, "Comparative analysis of soft computing techniques for predicting software effort based use case points," *IET Software*, vol. 12, no. 1. pp. 19–29, 2018.

[18] R. Silhavy, P. Silhavy, and Z. Prokopova, "Evaluating subset selection methods for use case points estimation," *Inf. Softw. Technol.*, vol. 97, pp. 1–9, May 2018.

[19] M. Azzeh and A. B. Nassif, "Analyzing the relationship between project productivity and environment factors in the use case points method," *J. Softw. Evol. Process*, vol. 29, no. 9, 2017.

[20] M. Azzeh and A. B. Nassif, "Project productivity evaluation in early software effort estimation," *J. Softw. Evol. Process*, vol. 30, no. 12, p. e2110, 2018.

[21] S. Diev, "Use cases modeling and software estimation: applying use case points," *SIGSOFT Softw.Eng.Notes*, vol. 31, no. 6, pp. 1–4, 2006.

[22] B. Anda, H. Dreiem, D. I. K. Sjøberg, and M. Jørgensen, "Estimating software development effort based on use cases – experiences from industry," in *Lecture Notes in Computer Science (including subseries Lecture Notes in Artificial Intelligence and Lecture Notes in Bioinformatics)*, 2001.

[23] K. Periyasamy and A. Ghode, "Cost Estimation Using Extended Use Case Point (e-UCP) Model," in *International Conference on Computational Intelligence and Software Engineering*, 2009.

[24] F. Wang, X. Yang, X. Zhu, and L. Chen, "Extended Use Case Points Method for Software Cost Estimation," in *International Conference on Computational Intelligence and Software Engineering*, 2009.

[25] M. R. Braz and S. R. Vergilio, "Software Effort Estimation Based on Use Cases," *30th Annual International Computer Software and Applications Conference COMPSAC06*. 2006.

[26] A. B. Nassif, L. Fernando Capretz, and D. Ho, "Estimating Software Effort Based on Use Case Point Model Using Sugeno Fuzzy Inference System," *2011 IEEE 23rd Int. Conf. Tools with Artif. Intell.*, 2011.

[27] G. Robiolo, C. Badano, and R. Orosco, "Transactions and paths: Two use case based metrics which improve the early effort estimation," in *International Symposium on Empirical Software Engineering and Measurement*, 2009, vol. 0, pp. 422–425.

[28] G. Robiolo and R. Orosco, "Employing use cases to early estimate effort with simpler metrics," *Innov. Syst. Softw. Eng.*, vol. 4, no. 1, pp. 31–43, 2008.

[29] L. Lavazza and G. Robiolo, "The role of the measure of functional complexity in effort estimation," in *ACM International Conference Proceeding Series*, 2010.

[30] M. Ochodek, J. N.-F. of C. and Decision, and U. 2010, "Enhancing use-case-based effort estimation with transaction types," *Found. Comput. Decis. Sci.*, vol. 35, no. 2, 2010.

[31] Sarwosri, M. J. Al Haiyan, M. Husein, and A. Putra Ferza, "The development of method of the enhancement of Technical Factor (TF) and Environmental Factor (EF) to the Use Case Point (UCP) to calculate the estimation of software's effort," in *Proceedings of 2016 International Conference on Information and Communication Technology and Systems, ICTS 2016*, 2017.

[32] M. Azzeh, A. Bou Nassif, S. Banitaan, and C. Lopez-Martin, "Ensemble of Learning Project Productivity in Software Effort Based on Use Case Points," in *Proceedings - 17th IEEE International Conference on Machine Learning and Applications, ICMLA 2018*, 2019.

[33] B. Anda, "Comparing Effort Estimates Based on Use Case Points with Expert Estimates," in *Empirical Assessment in Software Engineering*, 2002, pp. 8–10.

[34] B. Anda, E. Angelvik, and K. Ribu, "Improving estimation practices by applying use case models," in *Product Focused Software Process Improvement*, Springer, 2002, pp. 383–397.

[35] M. R. Braz and S. R. Vergilio, "Using fuzzy theory for effort estimation of object-oriented software," in *Proceedings - International Conference on Tools with Artificial Intelligence, ICTAI*, 2004.

[36] S. Kusumoto, F. Matukawa, K. Inoue, S. Hanabusa, and Y. Maegawa, "Estimating effort by use case points: Method, tool and case study," in *Proceedings - International Software Metrics Symposium*, 2004.

[37] B. Anda, H. C. Benestad, and S. E. Hove, "A multiple-case study of software effort estimation based on use case points," in *2005 International Symposium on Empirical Software Engineering, ISESE 2005*, 2005.

[38] E. R. Carroll, "Estimating software based on use case points," in *Companion to the 20th annual ACM SIGPLAN conference on Object-oriented programming, systems, languages, and applications*, 2005,

pp. 257–265.

[39] C. Gencel, L. Buglione, O. Demirors, and P. Efe, “A Case Study on the Evaluation of COSMIC-FFP and Use Case Points,” in *Software Measurement European Forum*, 2006, vol. Italy, pp. 121–140.

[40] R. K. Clemmons, “Project estimation with Use Case Points,” *J. Def. Softw. Eng.*, vol. 19, no. 2, 2006.

[41] M. Ochodek, B. Alchimowicz, ... J. J.-I. and S., and U. 2011, “Improving the reliability of transaction identification in use cases,” *Inf. Softw. Technol.*, vol. 53, no. 8, pp. 885–897, 2011.

[42] S. Frohnhoff and G. Engels, “Revised Use Case Point Method-Effort Estimation in Development Projects for Business Applications,” in *11th International Conference on Quality Engineering in Software Technology (CONQUEST 2008)*, 2008.

[43] R. Palucci Pantoni, E. A. Mossin, and D. Brandão, “Task Effort Fuzzy Estimator for Software Development,” *INFOCOMP*, vol. 7, no. 2, 2008.

[44] S. Ajitha, T. V. S. Kumar, D. E. Geetha, and K. R. Kanth, “Neural network model for software size estimation using use case point approach,” in *International Conference on Industrial and Information Systems (ICIIS)*, 2010, vol. Mangalore, pp. 372–376.

[45] G. B. Ibarra and P. Vilain, “Software estimation based on use case size,” in *Proceedings - 24th Brazilian Symposium on Software Engineering, SBES 2010*, 2010.

[46] A. B. Nassif, L. Capretz, and D. H. Emerging, “Software estimation in the early stages of the software life cycle,” in *International conference on emerging trends in computer science, communication and information technology*, 2010, pp. 5–13.

[47] A. B. Nassif, L. F. Capretz, and D. Ho, “A Regression Model with Mamdani Fuzzy Inference System for Early Software Effort Estimation Based on Use Case Diagrams,” in *Third International Conference on Intelligent Computing and Intelligent Systems*, 2011, vol. Guangzhou, pp. 615–620.

[48] Q. Yu, C. Liu, N. Li, and N. Ji, “Application of estimating based on use cases in Software Industry,” in *7th International Conference on Wireless Communications, Networking and Mobile Computing, WiCOM 2011*, 2011.

[49] S. Dash and A. A. Acharya, “Cost estimation for distributed systems using synthesized use case point model,” in *Communications in Computer and Information Science*, 2011.

[50] J. Lee, W. T. Lee, and J. Y. Kuo, “Fuzzy logic as a basic for use case point estimation,” in *IEEE International Conference on Fuzzy Systems*, 2011.

[51] N. Nunes, L. Constantine, and R. Kazman, “IUCP: Estimating interactive-software project size with enhanced use-case points,” *IEEE Softw.*, 2011.

[52] A. B. Nassif, D. Ho, and L. F. Capretz, “Regression Model for Software Effort Estimation Based on the Use Case Point Method,” in *2011 International Conference on Computer and Software Modeling*, 2011, vol. Singapore, pp. 117–121.

[53] Y. Yavari, M. Afsharchi, and M. Karami, “Software complexity level determination using software effort estimation use case points metrics,” in *2011 5th Malaysian Conference in Software Engineering, MySEC 2011*, 2011.

[54] C. Arumugam and C. Babu, “Developmental Size Estimation for Object-Oriented Software Based on Analysis Model,” *Int. J. Softw. Eng. Knowl. Eng.*, vol. 23, no. 3, pp. 289–308, 2013.

[55] J. Alwidian and W. Hadi, "Enhancing the results of UCP in cost estimation using new external environmental factors," *2012 Int. Conf. Inf. Technol. e-Services, ICITeS 2012*, 2012.

[56] A. B. Nassif, L. F. Capretz, and D. Ho, "Estimating software effort using an ANN model based on use case points," in *Proceedings - 2012 11th International Conference on Machine Learning and Applications, ICMLA 2012*, 2012, vol. 2, pp. 42–47.

[57] A. B. Nassif, L. F. Capretz, and D. Ho, "Software effort estimation in the early stages of the software life cycle using a cascade correlation neural network model," in *Proceedings - 13th ACIS International Conference on Software Engineering, Artificial Intelligence, Networking, and Parallel/Distributed Computing, SNPD 2012*, 2012.

[58] Z. C. Ani and S. Basri, "A CASE STUDY OF EFFORT ESTIMATION IN AGILE SOFTWARE DEVELOPMENT USING USE CASE POINTS," *Malaysia. Sci.Int.(Lahore)*, vol. 25, no. 4, pp. 1111–1126, 2013.

[59] L. M. Alves, A. Sousa, P. Ribeiro, and R. J. Machado, "An empirical study on the estimation of software development effort with use case points," in *Proceedings - Frontiers in Education Conference, FIE*, 2013.

[60] T. E. Ayyıldız, "Comparison of Three Software Effort Estimation Methodologies with Case Study," *AWERProcedia Inf. Technol. Comput. Sci.*, vol. 04, pp. 257–262, 2013.

[61] A. W. M. M. Parvez, "Efficiency factor and risk factor based user case point test effort estimation model compatible with agile software development," in *Proceedings - 2013 International Conference on Information Technology and Electrical Engineering: "Intelligent and Green Technologies for Sustainable Development", ICITEE 2013*, 2013.

[62] R. Alves, P. Valente, and N. J. Nunes, "Improving software effort estimation with human-centric models: a comparison of UCP and iUCP accuracy," in *Proceedings of the 5th ACM SIGCHI symposium on Engineering interactive computing systems*, 2013, pp. 287–296.

[63] C. Nagar and A. Dixit, "Efforts Estimation by Use Case Point using Experience Data," *Int. J. Comput. Appl.*, vol. 61, no. 17, pp. 975–8887, 2013.

[64] M. Azzeh, "Software cost estimation based on use case points for global software development," in *2013 5th International Conference on Computer Science and Information Technology, CSIT 2013 - Proceedings*, 2013.

[65] N. A. Ahmed and A. H. Ahmed, "Enabling complexity use case function point on service-oriented architecture," in *Proceedings - 2013 International Conference on Computer, Electrical and Electronics Engineering: "Research Makes a Difference", ICCEEE 2013*, 2013.

[66] A. B. Nassif, L. F. Capretz, and D. Ho, "Calibrating use case points," in *36th International Conference on Software Engineering, ICSE Companion 2014 - Proceedings*, 2014.

[67] S. M. Satapathy and S. K. Rath, "Use Case Point Approach Based Software Effort Estimation using Various Support Vector Regression Kernel Methods," *arXiv Prepr. arXiv1401.3069*, vol. 1, no. 1, pp. 1–13, 2014.

[68] P. Jovan, P. Sofija, B. M.-... F. T. (TELFOR), and U. 2015, "Enhancing use case point estimation method using fuzzy algorithms," in *23rd Telecommunications Forum Telfor (TELFOR)*, 2015.

[69] M. Saroha, S. S.-I. C. on Computing, and U. 2015, "Software effort estimation using enhanced use case point model," in *International Conference on Computing, Communication & Automation*, 2015.

[70] M. Azzeh, A. B. Nassif, and S. Banitaan, "An application of classification and class decomposition to use case point estimation method," in *Proceedings - 2015 IEEE 14th International Conference on Machine Learning and Applications, ICMLA 2015*, 2016, pp. 1268–1271.

[71] Y. Xie, J. Guo, and A. Shen, "Use case points method of software size measurement based on fuzzy inference," in *Lecture Notes in Electrical Engineering*, 2015.

[72] S. K. Rath, B. P. Acharya, and S. M. Satapathy, "Early stage software effort estimation using random forest technique based on use case points," *IET Softw.*, vol. 10, no. 1, 2016.

[73] S. Khatri, S. Malhotra, P. J.-2016 5th International, and U. 2016, "Use case point estimation technique in software development," in *2016 5th international conference on reliability, infocom technologies and optimization (trends and future directions)(ICRITO*, 2016.

[74] K. Iskandar, F. L. Gaol, B. Soewito, H. L. H. S. Warnars, and R. Kosala, "Software size measurement of knowledge management portal with use case point," in *Proceeding - 2016 International Conference on Computer, Control, Informatics and its Applications: Recent Progress in Computer, Control, and Informatics for Data Science, IC3INA 2016*, 2017.

[75] B. K. Park, S. Y. Moon, and R. Y. C. Kim, "Improving Use Case Point (UCP) Based on Function Point (FP) Mechanism," in *2016 International Conference on Platform Technology and Service, PlatCon 2016 - Proceedings*, 2016.

[76] R. Silhavy, P. Silhavy, and Z. Prokopova, "Analysis and selection of a regression model for the Use Case Points method using a stepwise approach," *J. Syst. Softw.*, vol. 125, 2017.

[77] M. Badri, L. Badri, W. Flageol, and F. Toure, "Source code size prediction using use case metrics: an empirical comparison with use case points," *Innov. Syst. Softw. Eng.*, vol. 13, no. 2–3, 2017.

[78] H. Leslie Hendric Spits Warnars, E. Abdurachman, and F. Lumban Gaol, "Use case point as software size measurement with study case of Academic Information System."

[79] D. Kurniadi, S. Sasmoko, H. L. H. S. Warnars, and F. L. Gaol, "Software size measurement of student information terminal with use case point," in *2017 IEEE International Conference on Cybernetics and Computational Intelligence, CyberneticsCOM 2017 - Proceedings*, 2018.

[80] Z. Prokopova, R. Silhavy, and P. Silhavy, "The effects of clustering to software size estimation for the use case points methods," in *Advances in Intelligent Systems and Computing*, 2017, pp. 479–490.

[81] S. Bagheri and A. Shameli-Sendi, "Software Project Estimation Using Improved Use Case Point," in *2018 IEEE 16th International Conference on Software Engineering Research, Management and Applications (SERA)*, 2018.

[82] K. Qi, B. B.-P. of the 10th I. W. On, and U. 2018, "Detailed use case points (DUCPs): a size metric automatically countable from sequence and class diagrams," in *2018 IEEE/ACM 10th International Workshop on Modelling in Software Engineering (MiSE)*, 2018.

[83] H. L. T. K. Nhung, H. T. Hoc, and V. Van Hai, "A Review of Use Case-Based Development Effort Estimation Methods in the System Development Context," in *Proceedings of the Computational Methods in Systems and Software.*, 2019, pp. 484–499.

[84] A. Effendi, R. Setiawan, Z. R.-P. C. Science, and U. 2019, "Adjustment Factor for Use Case Point Software Effort Estimation (Study Case: Student Desk Portal)," in *Procedia Computer Science*, 2019, pp. 691–698.

Appendix A. Selected papers

| ID | | Title | Ref |
|---|---|---|---|
| S1 | | Resource Estimation for Objectory Projects | [10] |
| S2 | | Estimating Software Development Effort Based on Use Cases –Experiences from Industry | [22] |
| S3 | | Comparing Effort Estimates Based on Use Case Points with Expert Estimates | [33] |
| S4 | | Improving Estimation Practices by Applying Use Case Models | [34] |
| S5 | | Using Fuzzy Theory for Effort Estimation of Object-Oriented Software | [35] |
| S6 | | Estimating Effort by Use Case Points: Method, Tool and Case Study | [36] |
| S7 | | A Multiple Case Study of Effort Estimation Based on Use Case Points | [37] |
| S8 | | Effort Estimation of Use Cases for Incremental Large-Scale Software Development | [4] |
| S9 | | Estimating Software Based on Use Case Points | [38] |
| S10 | A1 | A Case Study on The Evaluation Of COSMIC-FFP And Use Case Points | [39] |
| S11 | | Project Estimation with Use Case Points | [40] |
| S12 | | Software Effort Estimation Based on Use Cases | [25] |
| S13 | | Improving the reliability of transaction identification in use cases | [41] |
| S14 | | Revised Use Case Point Method - Effort Estimation in Development Projects for Business Applications | [42] |
| S15 | A2 | Employing Use Cases to Early Estimate Effort with Simpler Metrics | [28] |
| S16 | | Task Effort Fuzzy Estimator for Software Development | [43] |
| S17 | A3 | Cost Estimation Using Extended Use Case Point (E-UCP) Model | [23] |
| S18 | | Extended Use Case Point Method for Software Cost Estimation | [24] |
| S19 | | Transactions and Paths: Two Use Case Based Metrics Which Improve the Early Effort Estimation | [27] |
| S20 | A4 | Neural Network Model for Software Size Estimation Using Use Case Point Approach | [44] |
| S21 | | Software Estimation Based on Use Case Size | [45] |
| S22 | | Software Estimation in The Early Stages of The Software Life Cycle | [46] |
| S23 | | The Role of The Measure of Functional Complexity in Effort Estimation | [29] |
| S24 | | Enhancing Use Case Points Estimation Method Using Soft Computing Techniques | [15] |
| S25 | | Enhancing Use-Case-Based Effort Estimation with Transaction Types | [30] |
| S26 | | A Regression Model with Mamdani Fuzzy Inference System for Early Software Effort Estimation Based on Use Case Diagrams | [47] |
| S27 | | Application of Estimating Based on Use Cases in Software Industry | [48] |
| S28 | | Cost Estimation for Distributed Systems Using Synthesized Use Case Point Model | [49] |
| S29 | A5 | Estimating Software Effort Based on Use Case Point Model Using Sugeno Fuzzy Inference System. | [26] |
| S30 | | Fuzzy Logic as A Basic for Use Case Point Estimation | [50] |
| S31 | | iUCP: Estimating Interactive-Software Project Size with Enhanced Use-Case Points | [51] |
| S32 | | Regression Model for Software Effort Estimation Based on The Use Case Point Method | [52] |
| S33 | | Software Complexity Level Determination Using Software Effort Estimation Use Case Points Metrics | [53] |
| S34 | | Developmental Size Estimation for Object-Oriented Software Based on Analysis Model | [54] |
| S35 | | Simplifying Effort Estimation Based on Use Case Points | [1] |
| S36 | | A Treeboost Model for Software Effort Estimation Based on Use Case Points | [5] |
| S37 | | Enhancing the Results of UCP In Cost Estimation Using New External Environmental Factors | [55] |
| S38 | | Estimating Software Effort Using an ANN Model Based on Use Case Points | [56] |
| S39 | A6 | Software Effort Estimation in The Early Stages of The Software Life Cycle Using A Cascade Correlation Neural Network Model | [57] |
| S40 | A7 | Object Oriented Software Effort Estimate with Adaptive Neuro Fuzzy Use Case Size Point (ANFUSP) | [16] |
| S41 | | Towards an Early Software Estimation Using Log-Linear Regression and A Multilayer Perceptron Model | [3] |
| S42 | | A Case Study of Effort Estimation In a Agile Software Development Using Use Case Points | [58] |
| S43 | | An Empirical Study on The Estimation of Software Development Effort with Use Case Points | [59] |
| S44 | A8 | Comparison of Three Software Effort Estimation Methodologies with Case Study | [60] |
| S45 | | Efficiency Factor and Risk Factor Based User Case Point Test Effort Estimation Model Compatible with Agile Software Development | [61] |
| S46 | | Improving Software Effort Estimation with Human-Centric Models: A Comparison of UCP and iUCP Accuracy | [62] |
| S47 | A9 | Efforts Estimation by Use Case Point Using Experience Data | [63] |
| S48 | | Software Cost Estimation Based on Use Case Points for Global Software Development | [64] |
| S49 | | Enabling Complexity Use Case Function Point on Service-Oriented Architecture | [65] |
| S50 | | Calibrating Use Case Points | [66] |
| S51 | | Use Case Point Approach Based Software Effort Estimation Using Various Support Vector Regression Kernel Methods | [67] |
| S52 | | Enhancing Use Case Point Estimation Method Using Fuzzy Algorithms | [68] |
| S53 | | Software Effort Estimation Using Enhanced Use Case Point Model | [69] |
| S54 | A10 | An Application of Classification and Class Decomposition to Use Case Point Estimation Method | [70] |
| S55 | | Use Case Points Method of Software Size Measurement Based on Fuzzy Inference | [71] |
| S56 | A11 | A Hybrid Model for Estimating Software Project Effort from Use Case Points | [2] |
| S57 | | Early Stage Software Effort Estimation Using Random Forest Technique Based on Use Case Points | [72] |
| S58 | | Use Case Point Estimation Technique in Software Development | [73] |
| S59 | | Software Size Measurement of Knowledge Management Portal with Use Case Point | [74] |
| S60 | A12 | Improving Use Case Point (UCP) Based on Function Point (FP) Mechanism | [75] |
| S61 | | The Development of Method of The Enhancement of Technical Factor (TF) And Environmental Factor (EF) to the Use Case Point (UCP) To Calculate The Estimation of Software's Effort | [31] |
| S62 | | Analyzing the Relationship Between Project Productivity and Environment Factors in The Use Case Points Method | [19] |
| S63 | A13 | Analysis and Selection of a Regression Model for The Use Case Points Method Using A Stepwise Approach | [76] |
| S64 | | Source Code Size Prediction Using Use Case Metrics: An Empirical Comparison with Use Case Points. | [77] |
| S65 | | Use Case Point as Software Size Measurement with Study Case of Academic Information System, | [78] |
| S66 | | Software Size Measurement of Student Information Terminal with Use Case Point | [79] |
| S67 | | The Effects of Clustering to Software Size Estimation for The Use Case Points Methods. | [80] |
| S68 | | Comparative Analysis of Soft Computing Techniques for Predicting Software Effort Based Use Case Points | [17] |
| S69 | | Project Productivity Evaluation in Early Software Effort Estimation | [20] |
| S70 | | Evaluating Subset Selection Methods for Use Case Points Estimation | [18] |
| S71 | | Software Project Estimation Using Improved Use Case Point | [81] |
| S72 | A14 | Detailed Use Case Points (Ducps): A Size Metric Automatically Countable from Sequence and Class Diagrams | [82] |
| S73 | | Ensemble of Learning Project Productivity in Software Effort Based on Use Case Points | [32] |
| S74 | | A Review of Use Case-Based Development Effort Estimation Methods in The System Development Context | [83] |
| S75 | A15 | Adjustment Factor for Use Case Point Software Effort Estimation (Study Case: Student Desk Portal) | [84] |

## Appendix B. Quality assessment results

| ID | QA1 | QA2 | QA3 | QA4 | QA5 | QA6 | QA7 | Total | ID | QA1 | QA2 | QA3 | QA4 | QA5 | QA6 | QA7 | Total |
|---|---|---|---|---|---|---|---|---|---|---|---|---|---|---|---|---|---|
| S1 | 1 | 1 | 1 | 1 | 0 | 0 | 0.5 | 4.5 | S39 | 1 | 1 | 1 | 1 | 1 | 0 | 0.5 | 5.5 |
| S2 | 1 | 1 | 1 | 1 | 0 | 1 | 0.5 | 5.5 | S40 | 1 | 1 | 1 | 1 | 0 | 0 | 0.5 | 4.5 |
| S3 | 1 | 1 | 1 | 1 | 0 | 0 | 1 | 5 | S41 | 1 | 1 | 1 | 1 | 1 | 1 | 1 | 7 |
| S4 | 1 | 1 | 1 | 1 | 0 | 0 | 0.5 | 4.5 | S42 | 1 | 1 | 1 | 1 | 0 | 0 | 0.5 | 4.5 |
| S5 | 1 | 1 | 1 | 1 | 0 | 0 | 1 | 5 | S43 | 1 | 1 | 1 | 1 | 0 | 0 | 0.5 | 4.5 |
| S6 | 1 | 1 | 1 | 1 | 0 | 1 | 0.5 | 5.5 | S44 | 1 | 1 | 1 | 1 | 0.5 | 0 | 0 | 4.5 |
| S7 | 1 | 1 | 1 | 1 | 0 | 0 | 1 | 5 | S45 | 1 | 1 | 1 | 1 | 0.5 | 0 | 0 | 4.5 |
| S8 | 1 | 1 | 1 | 1 | 0 | 0 | 1 | 5 | S46 | 1 | 1 | 1 | 1 | 1 | 0 | 0.5 | 5.5 |
| S9 | 1 | 1 | 1 | 1 | 1 | 0 | 0.5 | 5.5 | S47 | 1 | 1 | 1 | 1 | 0 | 0 | 0.5 | 4.5 |
| S10 | 1 | 1 | 1 | 1 | 0 | 0 | 0.5 | 4.5 | S48 | 1 | 1 | 1 | 1 | 0 | 0 | 0.5 | 4.5 |
| S11 | 1 | 1 | 1 | 1 | 0 | 0 | 0.5 | 4.5 | S49 | 1 | 1 | 1 | 1 | 0 | 0 | 0.5 | 4.5 |
| S12 | 1 | 0.5 | 1 | 1 | 0 | 0 | 1 | 4.5 | S50 | 1 | 1 | 1 | 0.5 | 0 | 0 | 1 | 4.5 |
| S13 | 1 | 1 | 1 | 1 | 0 | 0 | 1 | 5 | S51 | 1 | 1 | 1 | 1 | 0 | 0 | 0.5 | 4.5 |
| S14 | 1 | 1 | 1 | 1 | 0 | 0 | 0.5 | 4.5 | S52 | 1 | 1 | 1 | 1 | 0 | 0 | 0.5 | 4.5 |
| S15 | 1 | 1 | 1 | 1 | 0 | 1 | 0.5 | 5.5 | S53 | 1 | 1 | 1 | 1 | 0 | 0 | 0.5 | 4.5 |
| S16 | 1 | 1 | 1 | 1 | 0 | 0 | 0.5 | 4.5 | S54 | 1 | 1 | 1 | 1 | 1 | 0 | 0.5 | 5.5 |
| S17 | 1 | 1 | 1 | 1 | 0 | 0 | 0.5 | 4.5 | S55 | 1 | 1 | 1 | 1 | 0 | 0 | 0.5 | 4.5 |
| S18 | 1 | 1 | 1 | 1 | 0 | 0 | 0.5 | 4.5 | S56 | 1 | 1 | 1 | 1 | 1 | 1 | 1 | 7 |
| S19 | 1 | 1 | 1 | 1 | 0 | 0 | 0.5 | 4.5 | S57 | 1 | 1 | 1 | 1 | 0 | 0 | 1 | 5 |
| S20 | 1 | 1 | 1 | 1 | 0 | 0 | 0.5 | 4.5 | S58 | 1 | 1 | 1 | 1 | 0 | 0 | 0.5 | 4.5 |
| S21 | 1 | 1 | 1 | 1 | 0 | 0 | 0.5 | 4.5 | S59 | 1 | 1 | 1 | 1 | 0 | 0 | 0.5 | 4.5 |
| S22 | 1 | 1 | 1 | 1 | 0 | 0 | 0.5 | 4.5 | S60 | 1 | 1 | 1 | 1 | 0 | 0 | 0.5 | 4.5 |
| S23 | 1 | 1 | 1 | 1 | 0 | 1 | 0.5 | 5.5 | S61 | 1 | 1 | 1 | 1 | 0 | 0 | 0.5 | 4.5 |
| S24 | 1 | 1 | 1 | 1 | 0 | 1 | 0 | 5 | S62 | 1 | 1 | 1 | 1 | 1 | 1 | 0.5 | 6.5 |
| S25 | 1 | 1 | 1 | 1 | 0 | 1 | 0.5 | 5.5 | S63 | 1 | 1 | 1 | 1 | 1 | 1 | 1 | 7 |
| S26 | 1 | 1 | 1 | 1 | 0 | 0 | 0.5 | 4.5 | S64 | 1 | 1 | 1 | 1 | 1 | 1 | 1 | 7 |
| S27 | 1 | 1 | 1 | 1 | 0 | 0 | 0.5 | 4.5 | S65 | 1 | 1 | 1 | 1 | 0 | 0 | 0.5 | 4.5 |
| S28 | 1 | 1 | 1 | 1 | 0 | 0 | 0.5 | 4.5 | S66 | 1 | 1 | 1 | 1 | 0 | 0 | 0.5 | 4.5 |
| S29 | 1 | 1 | 1 | 1 | 0 | 0 | 1 | 5 | S67 | 1 | 1 | 1 | 1 | 0 | 0 | 0.5 | 4.5 |
| S30 | 1 | 1 | 1 | 1 | 0 | 0 | 0.5 | 4.5 | S68 | 1 | 1 | 1 | 1 | 1 | 1 | 0.5 | 6.5 |
| S31 | 0.5 | 1 | 1 | 1 | 0 | 0 | 1 | 4.5 | S69 | 1 | 1 | 1 | 1 | 0 | 0 | 0.5 | 4.5 |
| S32 | 1 | 1 | 1 | 1 | 0 | 0 | 0.5 | 4.5 | S70 | 1 | 1 | 1 | 1 | 1 | 1 | 1 | 7 |
| S33 | 1 | 1 | 1 | 1 | 0 | 0 | 0.5 | 4.5 | S71 | 1 | 1 | 1 | 1 | 0 | 0 | 0.5 | 4.5 |
| S34 | 1 | 1 | 1 | 1 | 1 | 0 | 0.5 | 5.5 | S72 | 1 | 1 | 1 | 1 | 0 | 1 | 0.5 | 5.5 |
| S35 | 1 | 1 | 1 | 1 | 1 | 1 | 1 | 7 | S73 | 1 | 1 | 1 | 1 | 0 | 0 | 0.5 | 4.5 |
| S36 | 1 | 1 | 1 | 1 | 1 | 1 | 0.5 | 6.5 | S74 | 1 | 1 | 1 | 1 | 0 | 1 | 0.5 | 5.5 |
| S37 | 1 | 1 | 1 | 1 | 0 | 0 | 0.5 | 4.5 | S75 | 1 | 1 | 1 | 1 | 0 | 1 | 0.5 | 5.5 |
| S38 | 1 | 1 | 1 | 1 | 1 | 1 | 0.5 | 6.5 | | | | | | | | | |

Appendix C. Classification of the selected studies.

| ID | Source Type | Year | Contribution Type | Research Approach | Techniques | Digital Library |
|---|---|---|---|---|---|---|
| S1 | white paper | 1993 | Technique | Proposal + Case Study | Kerner | Google Scholar |
| S2 | Conference | 2001 | Validation | Case Study | Kerner | Springer |
| S3 | Conference | 2002 | Comparison | Comparative studies | Kerner | Springer |
| S4 | Conference | 2002 | Enhancement | Case Study | Kerner | Springer |
| S5 | Conference | 2004 | Enhancement | Evaluation | Fuzzy Logic | IEEE |
| S6 | Conference | 2004 | Tool | Case Study | Kerner | IEEE |
| S7 | Conference | 2005 | Validation | Case Study | Kerner | IEEE |
| S8 | Conference | 2005 | Enhancement | Case Study | Kerner | ACM |
| S9 | Conference | 2005 | Validation | Case Study | Kerner | ACM |
| S10 | Conference | 2006 | Comparison | Case Study | Kerner | Google Scholar |
| S11 | Journal | 2006 | Validation | Case Study | Kerner | Google Scholar |
| S12 | Conference | 2006 | Technique | Evaluation + Comparative studies | Kerner | IEEE |
| S13 | Journal | 2011 | Model | Comparative studies | NN+SR | Science Direct |
| S14 | Conference | 2008 | Enhancement | Evaluation | Kerner | Google Scholar |
| S15 | Journal | 2008 | Technique | Case Study | Kerner | Springer |
| S16 | Journal | 2008 | Technique | Evaluation | Fuzzy Logic | Google Scholar |
| S17 | Conference | 2009 | Enhancement | Theory | Kerner | IEEE |
| S18 | Conference | 2009 | Model | Case Study | Fuzzy Logic + BBN | IEEE |
| S19 | Conference | 2009 | Enhancement | Evaluation + Case Study | Kerner | ACM |
| S20 | Conference | 2010 | Model | Case Study | NN | IEEE |
| S21 | Conference | 2010 | Technique | Evaluation | Kerner | IEEE |
| S22 | Conference | 2010 | Comparison | Comparative studies | Fuzzy Logic + Kerner | Google Scholar |
| S23 | Conference | 2010 | Validation | Evaluation | Fuzzy Logic + Kerner | ACM |
| S24 | Journal | 2010 | Technique | Evaluation | Fuzzy Logic + NN | Google Scholar |
| S25 | Journal | 2010 | Technique | Case Study | Kerner | Google Scholar |
| S26 | Conference | 2011 | Technique | Evaluation | Fuzzy Logic | IEEE |
| S27 | Conference | 2011 | Comparison | Comparative studies | Kerner | IEEE |
| S28 | Conference | 2011 | Enhancement | Theory | Kerner | Springer |
| S29 | Conference | 2011 | Technique | Evaluation | Fuzzy Logic | IEEE |
| S30 | Conference | 2011 | Technique | Evaluation | Fuzzy Logic | IEEE |
| S31 | Magazine | 2011 | Enhancement | Theory | Kerner | IEEE |
| S32 | Conference | 2011 | Technique | Evaluation | Regression | Google Scholar |
| S33 | Conference | 2011 | Validation | Evaluation | Kerner | IEEE |
| S34 | Journal | 2011 | Technique | Review + Survey | New Model | Google Scholar |
| S35 | Journal | 2011 | Validation | Evaluation | Karner + Regression | Science Direct |
| S36 | Conference | 2012 | Model | Evaluation | Treeboost + Regression | IEEE |
| S37 | Conference | 2012 | Enhancement | Proposal | Kerner | IEEE |
| S38 | Conference | 2012 | Model | Evaluation | ANN | IEEE |
| S39 | Conference | 2012 | Model | Evaluation | ANN | IEEE |
| S40 | Journal | 2012 | Model | Evaluation | ANFIS | Google Scholar |
| S41 | Journal | 2012 | Model | Evaluation | Regression | Science Direct |
| S42 | Conference | 2013 | Enhancement | Theory | N/A | Google Scholar |
| S43 | Conference | 2013 | Validation | Review | Kerner | IEEE |
| S44 | Conference | 2013 | Comparison | Comparative studies | Kerner | Google Scholar |
| S45 | Conference | 2013 | Technique | Evaluation | N/A | IEEE |
| S46 | Conference | 2013 | Validation + Model | Evaluation + Comparative studies | Kerner | ACM |
| S47 | Journal | 2013 | Enhancement | Evaluation | Kerner | Google Scholar |
| S48 | Conference | 2013 | Enhancement | Proposal | Kerner | IEEE |
| S49 | Conference | 2013 | Technique | Proposal + Case Study | Kerner | IEEE |
| S50 | Conference | 2014 | Enhancement | Comparative studies | Fuzzy Logic + NN | ACM |
| S51 | Journal | 2014 | Model | Comparative studies | SVM | Google Scholar |
| S52 | Conference | 2015 | Enhancement | Comparative studies | Fuzzy Logic | IEEE |
| S53 | Conference | 2015 | Enhancement | Case Study | Kerner | IEEE |
| S54 | Conference | 2015 | Technique | Evaluation + Comparative studies | Kerner | IEEE |
| S55 | Conference | 2015 | Enhancement | Evaluation | Kerner | Springer |
| S56 | Journal | 2016 | Model | Comparative studies | Kerner | Science Direct |
| S57 | Journal | 2016 | Model | Evaluation + Comparative studies | Kerner | IET Software |
| S58 | Conference | 2016 | Comparison | Proposal | N/A | IEEE |
| S59 | Conference | 2016 | Validation | Proposal | N/A | IEEE |
| S60 | Conference | 2016 | Enhancement | Evaluation | N/A | IEEE |
| S61 | Conference | 2016 | Enhancement | Evaluation | Kerner | IEEE |
| S62 | Journal | 2017 | Validation | Case Study | Kerner | Wiley |
| S63 | Journal | 2017 | Technique | Evaluation | Stepwise multiple linear regression | Science Direct |
| S64 | Journal | 2017 | Validation | Evaluation | simple linear regression | Springer |
| S65 | Conference | 2017 | Validation | Case Study + Evaluation | Kerner | IEEE |
| S66 | Conference | 2017 | Validation | Review | N/A | IEEE |
| S67 | Conference | 2017 | Validation | Evaluation | Kerner | Springer |
| S68 | Journal | 2018 | Comparison | Comparative studies | Kerner | IET Software |
| S69 | Journal | 2018 | Validation | Evaluation | Kerner | Wiley |
| S70 | Journal | 2018 | Validation | Evaluation | regression + subset selection | Science Direct |
| S71 | Conference | 2018 | Technique + Enhancement | Case Study | Kerner | IEEE |
| S72 | Conference | 2018 | Enhancement | Proposal + Case Study | Kerner | IEEE |
| S73 | Conference | 2018 | Technique | Comparative studies | Kerner | IEEE |
| S74 | Conference | 2019 | Validation | Review | N/A | Springer |
| S75 | Conference | 2019 | Validation | Case Study | Kerner | Science Direct |